\documentclass[jcp,amsmath,amssymb,amsfonts,reprint,graphicx,natbib, showpacs, showkeywords]{revtex4-1}
\usepackage{natbib} 
\usepackage{mathtools}
\usepackage{color}
\usepackage{wrapfig}

\draft % marks overfull lines with a black rule on the right

\newcommand{\rb}{\mathbf{r}}%
\newcommand{\dr}{\text{d}\mathbf{r}}%
\newcommand{\dt}{\partial_t}%

\newcommand{\eqdef}{\ensuremath{\overset{\text{def}}{=}}}

\newcommand{\rrho}{\Bar{\Bar{\varrho}}}

\begin{document}

% Use the \preprint command to place your local institutional report number 
% on the title page in preprint mode.
% Multiple \preprint commands are allowed.
%\preprint{}

\title[Navier-Stokes PFC model]{A Navier Stokes Phase Field Crystal Model for Colloidal Suspensions} %Title of paper

% repeat the \author .. \affiliation  etc. as needed
% \email, \thanks, \homepage, \altaffiliation all apply to the current author.
% Explanatory text should go in the []'s, 
% actual e-mail address or url should go in the {}'s for \email and \homepage.
% Please use the appropriate macro for the type of information

% \affiliation command applies to all authors since the last \affiliation command. 
% The \affiliation command should follow the other information.

\author{Simon Praetorius}
\email[]{simon.praetorius@tu-dresden.de}

\author{Axel Voigt}
\email[]{axel.voigt@tu-dresden.de}
%\homepage[]{Your web page}
%\thanks{}
%\altaffiliation{}
\affiliation{Institute of Scientific Computing, Technische Universit\"{a}t Dresden, D-01062 Dresden, Germany}

% Collaboration name, if desired (requires use of superscriptaddress option in \documentclass). 
% \noaffiliation is required (may also be used with the \author command).
%\collaboration{}
%\noaffiliation

\date{\today}

\begin{abstract}
We develop a fully continuous model for colloidal suspensions with hydrodynamic interactions. The Navier Stokes Phase Field Crystal (NS-PFC) model combines ideas of dynamic density functional theory with particulate flow approaches and is derived in detail and related to other dynamic density functional theory approaches with hydrodynamic interactions. The derived system is  numerically solved using adaptive finite elements and used to analyse colloidal crystallization in flowing environments demonstrating a strong coupling in both directions between the crystal shape and the flow field. We further validate the model against other computational approaches for particulate flow systems for various colloidal sedimentation problems.
\end{abstract}

% insert suggested PACS numbers in braces on next line

\maketitle %\maketitle must follow title, authors, abstract and \pacs

% If in two-column mode, this environment will change to single-column format so that long equations can be displayed. 
% Use only when necessary.
%\begin{widetext}
%$$\mbox{put long equation here}$$
%\end{widetext}

\section{\label{seq:introduction}Introduction}
Simple fluids can be coarse grained, considered as a continuum and very well described by the Navier-Stokes equations. A quantitative description can be achieved down to the nanometer scale. This simple treatment is not necessarily valid any more for colloidal suspensions. Here, colloidal particles with typical size of nanometers to a few microns move due to collisions with the solvent molecules, interact with each other and induce flow fields due to their motion. These hydrodynamic interactions are shown to be of relevance in various practical application, e.g. colloidal gelation \cite{Furukawa2010} or coagulation of colloidal dispersions \cite{Matsuoka2012}. To calculate nonequilibrium properties of such systems requires to resolve the different time- and length scales arising from thermal Brownian motion and hydrodynamic interactions. Various approaches have been developed to consider these interactions in an effective way. For an overview and proposed coarse-graining descriptions see e.g. \cite{Padding2006}.  One of the most popular approaches is Stokesian dynamics (SD) within the low Reynolds number limit \cite{Brady1988}. The hydrodynamic interaction is thereby incorporated in an approximate analytical form, assuming to result as the sum of two-body interactions. The approach is difficult to implement for complex boundary conditions and is relatively expensive. As an alternative, direct numerical simulations, which involve determining fluid motion simultaneously with particle motion, are proposed. In these methods, the colloidal particles are fully resolved and coupled with the Navier-Stokes equations, leading to coupled discrete-continuous descriptions. 

Our aim is to derive from these models a fully continuous system of equations. This has the advantage of an efficient numerical treatment, the possibility of a detailed numerical analysis and offers a straight forward coupling with other fields. The model will serve as a general continuum model for colloidal suspension, providing a quantitative approach down to the length scale set by the colloidal particles and is operating on diffusive time scales. The approach will be derived by combining ideas from: (a) dynamic density functional theory (DDFT), and (b) classical particulate flow systems. We will test the derived system for colloidal crystallization in flowing environments and for colloidal sedimentation. 

% ---------------------------------------------------------------------------------------------------
\subsection{Dynamic density functional theory approach}
The aim of the dynamic density functional theory (DDFT) approach is to provide a reduced model that describes the local state of a colloidal fluid by the time averaged one-particle density. The evolution of this density is driven by a gradient-flow of the equilibrium Helmholtz free-energy functional. The first realization of a DDFT for colloidal fluids is the work of Marconi and Tarazona \cite{Marconi2000} with colloids modelled as Brownian particles. Later this theory is extended by Archer \cite{Archer2009} and connected to the equations of motion from continuum fluid mechanics. Rauscher \cite{Rauscher2007} described an advected DDFT, to model colloids in a flowing environment, that do not interact via hydrodynamic interactions. The work of Goddard et.al. \cite{Goddard2013} incorporates the effect of inertia and hydrodynamic interactions between the colloidal particles and recently Gr\'an\'asy et.al. \cite{Toth2014} explored a coarse-grained density coupling of DDFT and the Navier-Stokes equations.

We start the derivation of our model with the dynamical equations derived by Archer \cite{Archer2009}. Therefore we introduce the one-body (number) density $\varrho(\mathbf{r},t)$ and the average local velocity $\mathbf{v}(\mathbf{r}, t)$ of the colloidal particles. The density is driven by a continuity equation
\begin{equation}\label{eq:continuity_equation}
\partial_t \varrho + \nabla\cdot(\varrho\mathbf{v}) = 0
\end{equation}
with the current, expressed as $\varrho\mathbf{v}$, evolving by the dynamical equation
\begin{equation}\label{eq:archer_velocity}
m\varrho\left(\partial_t\mathbf{v} + (\mathbf{v}\cdot\nabla)\mathbf{v} + \gamma\mathbf{v}\right) = -\varrho\nabla\frac{\delta\mathcal{F}_\text{H}[\varrho]}{\delta\varrho} + \eta\Delta\mathbf{v},
\end{equation}
where $m$ represents the mass of the particles, $\gamma$ a dumping coefficient, $\mathcal{F}_\text{H}[\varrho]$ the equilibrium Helmholtz free-energy functional and $\eta$ a viscosity coefficient.

We use a minimal expression for the free-energy, the Swift-Hohenberg (SH) energy \cite{Swift1977, Elder2002}, in dimensionless form
\begin{equation}\label{eq:swift-hohenberg-energy}
\mathcal{F}_\text{H}[\varrho(\psi)] \simeq \mathcal{F}_\text{sh}[\psi] = \int \frac{1}{4}\psi^4 + \frac{1}{2}\psi(r + (q_0^2+\Delta)^2)\psi\,\text{d}\hat{\mathbf{r}},
\end{equation}
with $\varrho = \bar{\varrho}( 1 + (\psi + 0.5) )$ a parametrization of the one-particle density with respect to a reference density $\bar{\varrho}$. The phenomenological parameter $r$ is related to the undercooling of the system and the constant $q_0$ is related to the lattice spacing. This functional arises by splitting the energy in an ideal gas contribution and an excess free energy $\mathcal{F}_\text{H} = \mathcal{F}_\text{id} + \mathcal{F}_\text{exc}$, rescaling and shifting of the order-parameter $\varrho$, expanding ideal gas contributions in real-space, and the excess free energy in Fourier-space and simplification by removing constant and linear terms that would vanish in the dynamical equations. A detailed derivation of the energy can be found in \cite{Teeffelen2009} upon others.

Inserting the density expansion and the free-energy \eqref{eq:swift-hohenberg-energy} into \eqref{eq:continuity_equation} and \eqref{eq:archer_velocity} we get a system of dynamic equations for the density deviation $\psi$ and the related non-dimensionalized averaged velocity $\hat{\mathbf{v}}$:
\begin{align}
\partial_t \psi + \nabla\cdot\big((1.5 + \psi)\hat{\mathbf{v}}\big) = &\;0 \label{eq:archer_velocity_non-dimensionlized} \\
(1.5 + \psi)\left(\partial_t\hat{\mathbf{v}} + (\hat{\mathbf{v}}\cdot\nabla)\hat{\mathbf{v}} + \Gamma\hat{\mathbf{v}}\right) = &\frac{1}{Re}\Delta\hat{\mathbf{v}} \notag \\
-\frac{1}{Pe}(1.5 + \psi)&\nabla\frac{\delta\mathcal{F}_\text{sh}[\psi]}{\delta\psi}. \label{eq:archer_velocity_non-dimensionlized2}
\end{align}
With respect to a length-scale $L$ and time-scale $L/V_0$ we have the dimensionless variable $\hat{\mathbf{v}} = \mathbf{v}/V_0$ and Peclet number $Pe$, Reynolds number $Re$ and friction coefficient $\Gamma$ given by
\[Pe = \frac{3m V_0^2}{k_\text{B} T},\quad Re = \frac{m\bar{\varrho} L V_0}{\eta},\quad \Gamma=\frac{\gamma L}{V_0},\]
with Boltzmann's constant $k_\text{B}$ and temperature $T$.
In the Appendix \ref{app:a} a detailed derivation of this dimensionless form of the dynamical equations can be found.

In the overdamped limit, $\Gamma \gg 1$, the velocity equation reduces to an explicit expression that relates the velocity to the chemical potential by
\begin{equation}\label{eq:limit_gamma_1}
(1.5+\psi)\hat{\mathbf{v}} = - \frac{1}{\Gamma Pe} (1.5 + \psi)\nabla\frac{\delta\mathcal{F}_\text{sh}[\psi]}{\delta\psi}.
\end{equation}
Inserting \eqref{eq:limit_gamma_1} into \eqref{eq:archer_velocity_non-dimensionlized} results in the PFC equation \cite{Elder2002}
\begin{equation}\label{eq:classical_pfc}
\partial_t \psi = \frac{1}{\Gamma Pe} \nabla\cdot\left((1.5 + \psi)\nabla\frac{\delta\mathcal{F}_\text{sh}[\psi]}{\delta\psi}\right),
\end{equation}
referred to as PFC1 model in \cite{Teeffelen2009}.

% ---------------------------------------------------------------------------------------------------
\subsection{Particulate flows}
Typical approaches to simulate particulate flows on larger length scales consider a  Newton-Euler equation for each particle to describe their motion as a rigid body and combine this with a Navier-Stokes solver for the flow around these particles. Various numerical approaches have been proposed to model this flow and the incorporation of a no-slip boundary condition on the particle surface, see e.g. \cite{Glowinski2001,Uhlmann2005,Apte2009,Kempe2012}.
Examples for numerical approaches are the fictitious domain and immersed boundary method. All these approaches use the general idea to consider the particles as a highly viscous fluid, which allows the flow computation to be done on a fixed space region.
The no-slip boundary condition on the particle surface is thereby enforced directly or implicitly, depending on the numerical approach. All these methods combine a continuous description of the flow field with a discrete off-lattice simulation for the particles.

Considering an incompressible fluid with viscosity $\eta_f$ and constant fluid density $\rho_f$, we can write the Navier-Stokes equations for velocity $\mathbf{u}$ and pressure $p$ of a pure fluid in dimensionless form:
\begin{align}
\partial_t\hat{\mathbf{u}} + (\hat{\mathbf{u}}\cdot\nabla)\hat{\mathbf{u}} &= -\nabla \hat{p} + \frac{1}{Re_f}\nabla\cdot\big(2(1+\tilde{\eta})\mathbf{D}(\hat{\mathbf{u}})\big) \label{eq:navier-stokes} \\
\nabla\cdot\hat{\mathbf{u}} &= 0, \label{eq:navier-stokes-div}
\end{align}
with length- and time-scale as above and dimensionless velocity $\hat{\mathbf{u}}=\mathbf{u}/V_0$, viscosity perturbation $\tilde{\eta}$ from the expansion $\eta_f = \bar{\eta}_f(1+\tilde{\eta})$, fluid Reynolds number $Re_f$ and dimensionless pressure $\hat{p}$ given by
\[Re_f = \frac{\rho_f L V_0}{\bar{\eta}_f},\quad \hat{p}=\frac{p}{\rho_f V_0^2},\]
respectively. The expression $\mathbf{D}(\hat{\mathbf{u}})$ gives the symmetric part of the velocity gradient, i.e.
\[\mathbf{D}(\hat{\mathbf{u}}) = \frac{1}{2}(\nabla\hat{\mathbf{u}} + \nabla\hat{\mathbf{u}}^\top).\]

As a reference model for colloidal suspensions we consider the fluid particle dynamics model (FPD) by \cite{Tanaka2000}. Here, the particles are considered as a highly viscous fluid and the velocities of the particles are extracted from the fluid velocity $\mathbf{u}$. The shape of the particles is constructed using a $\tanh$-profile with a specified radius and interface thickness and their centers of mass interact via an interparticle potential. The approach can also be seen as a modification of a classical ``Model H'' \cite{Siggia1976,Hohenberg1977}, with a fluid and a particle phase and the driving force in the Navier-Stokes equations governed by the interatomic potential. The approach again combines continuous and discrete descriptions. 

The motion of colloidal particles with positions $\mathbf{r}_i(t)$ are governed by the velocities $\mathbf{v}_i(t)$ and the evolution of a flow field $\mathbf{u}$, where the colloidal particles are suspended in. The basic idea is to introduce concentration fields $\phi_i(\mathbf{r},t)\in[0,1]$ for each particle and to average the fluid velocity over regions with high concentration, i.e. \[\mathbf{v}_i(t) = \frac{\int \phi_i(\mathbf{r},t)\mathbf{u}(\mathbf{r},t)\,\text{d}\mathbf{r}}{\int \phi_i(\mathbf{r},t)\,\text{d}\mathbf{r}}.\] Thus the motion of the particles can be described by $\rb_i(t+\Delta t) := \rb_i(t) + \Delta t\cdot\mathbf{v}_i(t)$, with $\Delta t$ the simulation time step.

A space-dependent fluid viscosity $\eta_f$, as a function of $\phi_i$, is introduced to describe the rigidity of the particles, and a force term $\mathbf{F}:=\mathbf{F}^{[\text{ta}]}$ to account for the particle interactions in the flow equation \eqref{eq:navier-stokes}. This force is chosen as the negative gradient of an interaction potential $\cal{V}$, multiplied with the particle-concentration fields $\phi_i$:
\begin{equation}\label{eq:force-term}
\mathbf{F}^{[\text{ta}]}(\mathbf{r}) \eqdef -\sum_i \nabla_{\rb_i} \big(\sum_{j\neq i} {\cal V}(\|\mathbf{r}_i - \mathbf{r}_j\|) \big) \phi_i(\mathbf{r}).
\end{equation}
The fluid viscosity $\eta_f = \bar{\eta}_f(1+\tilde{\eta})$ is modeled, by describing the viscosity perturbation $\tilde{\eta}$, as
\begin{equation}\label{eq:viscosity}
\tilde{\eta}(\mathbf{r})=\sum_i \big(\frac{\bar{\eta}_p}{\bar{\eta}_f}-1\big)\phi_i(\mathbf{r}),
\end{equation}
with $\bar{\eta}_f < \bar{\eta}_p$ the liquid and particle viscosity, respectively. In \cite{Nakayama2005} it is argued, that the artificial diffusivity $\bar{\eta}_p / \bar{\eta}_f$ must go to $\infty$ for the particles to become rigid. In their method, they have introduced a different body force to guarantee this rigidity without taking large values of the viscosity ratio. However, we will here only consider the original FPD approach.

% ---------------------------------------------------------------------------------------------------
\subsection{Towards a fully continuous description}
Our aim is to derive a fully continuous model by combining the FPD model with the PFC approach. A first step in this direction has already been done in \cite{Menzel2013}, where the interaction potential is already replaced by the PFC approach. The discrete off-lattice simulation for the particles is no longer needed, the particle positions and velocities result from the advected PFC model. However, the forcing term in the Navier-Stokes equations still requires to identify the position and velocity of each particle and thus the approach still has a discrete component. To derive a fully continuous model we will first clarify the relation of the different approaches in \cite{Praetorius2011,Tanaka2000,Menzel2013} and will show, that all the discrete coupling terms can be approximated with a simple continuous expression. 

To allow for a description of the flow of individual particles, we consider a variant of the PFC model, the vacancy PFC model, introduced in \cite{Chan2009, Berry2011}. Instead of minimizing the Swift-Hohenberg functional directly, we consider a density field with positive density deviation $\psi$, only, which leads to a modification of the particle-interaction and allows to handle single particles, as well as many individual particles embedded in the fluid.

%%%%%%%%%%%%%%%%%%%%%%%%%%%%%%%%%%%%%%%%%%%%%%%%%%%%%%%%%%%%%%%%%%%%%%%%%%%%%%%%%%%%%%%%%%%%%%%%%%%%%%%%%%%%%
\section{Derivation of a fully continuous model}
In \cite{Menzel2013} the PFC model and the FPD model are combined, by letting the density field influence the flow field. The interatomic potential is encoded in the Swift-Hohenberg energy \eqref{eq:swift-hohenberg-energy} and the particle positions evolve according the an advective PFC equation (see below). The forcing term $\mathbf{F}:=\mathbf{F}^{[\text{ml}]}$ in the Navier-Stokes equations now ensures the fluid velocity $\mathbf{u}$ to be equal to the particle velocity $\mathbf{v}_i$ at the particle position $\mathbf{r}_i$, i.e.
\begin{equation}\label{eq:force-term2}
\mathbf{F}^{[\text{ml}]}(\mathbf{r})\eqdef \omega\sum_i(\hat{\mathbf{v}}_i - \hat{\mathbf{u}}(\mathbf{r}))\delta(\mathbf{r} - \mathbf{r}_i),
\end{equation}
with $\omega\gg 1$ a penalty parameter and $\delta(\cdot)$ the pointwise delta-function. Thereby, position and velocity of each individual particle must be extracted from the density field $\psi$ by tracking the maxima of the density that are interpreted as average particle positions. These quantities are then explicitly inserted into the expression of the forcing term. The fluid viscosity $\eta_f$ can be modeled as before in \eqref{eq:viscosity}, but now $\psi$ can directly be used to distinguish between the background fluid and the particles. 

In the following, we give a new formulation of a continuous force term that can be evaluated without extracting individual particle positions and velocities. At first, we relate the density field $\psi$, described in \eqref{eq:classical_pfc}, to a delta function $\delta(\rb)$ and to a concentration field $\phi(\rb) = \sum_i \phi_i(\rb)$. In a second step, the particle velocities are shown to arise directly from the evolution equation \eqref{eq:classical_pfc}, respective \eqref{eq:limit_gamma_1}.

% ---------------------------------------------------------------------------------------------------
\subsection{Approximation of a delta-function}
%Let $\operatorname{dist}(\mathbf{r})$ be the distance function of $\mathbf{r}$ to the fixed center-of-mass coordinate of a particle located at the origin. The delta-function $\delta(\mathbf{r})$ can be approximated by
%\begin{equation}\label{eq:smeared-out-delta-function}
%\delta_\epsilon(\mathbf{r}):=\left\{\begin{array}{ll}
%\frac{1}{2\epsilon}\big(1+\cos(\frac{\pi\operatorname{dist}(\mathbf{r})}{\epsilon})\big) & \text{ for }|\operatorname{dist}(\mathbf{r})|\leq\epsilon \\
%0 & \text{ otherwise,}
%\end{array}\right. \!\!
%\end{equation}
%with $\epsilon>0$ a small parameter, that defines the width of the smeared out region of $\delta_\epsilon$, \cite[e.g.][]{Sussman1994,Engquist2005,Towers2007}. It can be shown, that $\delta_\epsilon$ converge weakly to $\delta$, i.e. $\delta_\epsilon(\mathbf{r})\xrightharpoonup{} \delta(\mathbf{r}) \;:\Leftrightarrow\;\langle \delta_\epsilon,\xi\rangle_{L_2}\xrightarrow{}\langle \delta,\xi\rangle_{L_2}$ as $\epsilon\rightarrow 0$, for all testfunctions $\xi$. Our aim is to derive a similar approximation using the density field $\psi$. 

\begin{figure}
\begin{center}
\includegraphics[width=.9\linewidth]{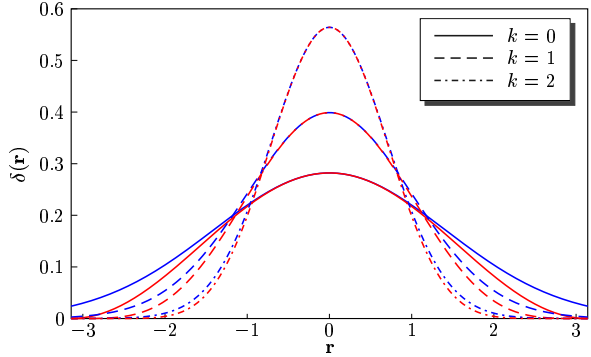}
\caption{(Color online) The first three elements of the sequences $\delta_{\epsilon_k}^\text{exp}$ (normalized), with $\epsilon_k = 2^{-k}$, in blue (upper curves) and $\delta_{(k)}$ in red (lower curves). The lattice constant is $q_0=1$.}\label{fig:nascent-delta-function}
\end{center}
\end{figure}

For the classical PFC equation in 1D, a one-mode approximation of the density $\psi$ is given by \cite{Elder2004}
\begin{equation}\label{eq:om-density}
\psi_\text{om}(\mathbf{r})=A\cos(q_0\mathbf{r})+\bar\psi,
\end{equation}
where $A, q_0$ and $\bar\psi$ are constants, that define the amplitude, lattice constant and mean density of the field, respectively. We introduce 
\begin{equation}\label{eq:psi-recursion}
\psi_{(0)} = \frac{1}{2}\Big(1+\frac{\psi_\text{om} - \bar\psi}{A}\Big), \quad
\psi_{(k)} = (\psi_{(k-1)})^{2},
\end{equation}
for $k>0$, or in explicit form $\psi_{(k)}=[\psi_{(0)}]^{2^k}$ for $k\in\mathbb{N}$. After appropriate normalization, we obtain 
\begin{equation}\label{eq:nascent-delta-function}
\delta_{(k)}(\mathbf{r}):=N_k\psi_{(k)}(\rb),
\end{equation}
with $N_k$ normalization constants, that ensure the property $\int\delta_{(k)}(\mathbf{r})\,\dr=1$. Values for various indices $k$ can be found in Table \ref{tbl:normalization_constants}. Thus we have a sequence of nascent delta functions. Figure \ref{fig:nascent-delta-function} shows the first three elements of this sequence in comparison with the classical Gaussians $\delta^\text{exp}_\epsilon(\rb)\cong e^{[(q_0\|\rb\|)^2/(-4\epsilon)]}$, visualizing the convergence qualitatively. As a consequence of this property, the shifted and scaled density field $\psi_{(0)}$ can be seen as a first-order approximation of a delta function. The approach can be generalized to 2D and 3D and will be used for $\psi$ instead of $\psi_{om}$.

\begin{table}
\begin{tabular}{c|c|c|c|c}
$k$ & $0$ & $1$ & $2$ & $3$ \\ 
\hline 
$N_k\cdot\frac{\pi}{q_0}$ & $1$ & $\frac{4}{3}$ & $\frac{64}{35}$ & $\frac{16384}{6435}$
\end{tabular}
\caption{The first four elements of the sequence $N_k$, the normalization constants for the nascent delta function $\delta_{(k)}$.}\label{tbl:normalization_constants}
\end{table}

% ---------------------------------------------------------------------------------------------------
\subsection{Approximation of concentration fields}
\begin{figure}
\begin{center}
\includegraphics[width=.9\linewidth]{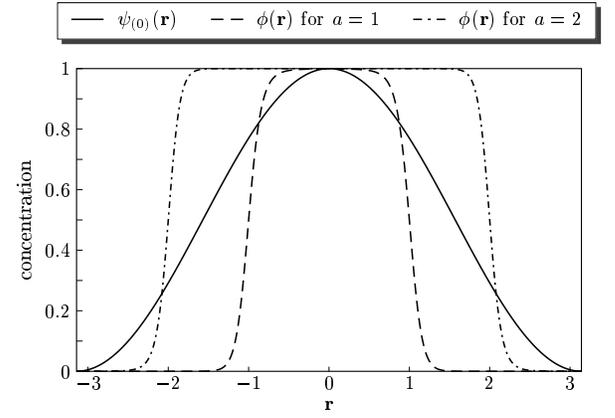}
\caption{Transformation of the density field $\psi$ into a $\operatorname{tanh}$-concentration field, for different particle radii. The lattice constant is $q_0=1$ and interface width $\epsilon=0.1$.}\label{fig:concentration-profile}
\end{center}
\end{figure}

The concentration field $\phi_i$ in \cite{Tanaka2000}, used for the phase-field description of particles, is defined by
\[\phi_i(\rb)=\frac{1}{2}\Big(1-\operatorname{tanh}\big((\|\rb-\rb_i\|-a) \frac{3}{\epsilon}\big)\Big),\]
with $\rb_i$ the center-of-mass position of the $i$th particle, $a$ the particle radius and $\epsilon$ a small parameter, that defines the width of the smoothing region. We now interpret $\psi_{(0)}$ in \eqref{eq:psi-recursion} as a concentration field. It has value one at the maxima of the cosine profile and zero in between. The transition is very coarse, but gives an approximation of the $\tanh$-profile of $\phi(\rb)=\sum_i\phi_i(\rb)$, which can be refined with
\begin{equation}\label{eq:concentration-field}
\phi(\psi)=\frac{1}{2}\Big(1+\operatorname{tanh}\big((\psi_{(0)}-\sigma)\frac{3}{\epsilon}\big)\Big),
\end{equation}
where $\sigma=\frac{1}{2}\big(1+\cos(q_0\cdot a)\big)$ is a shifting parameter, see Figure \ref{fig:concentration-profile} for a realization.

In order to define the viscosity field, we adopt the expression \eqref{eq:viscosity} and insert for $\phi(\rb)$ the field $\phi(\psi)$. Thus, we introduce a viscosity field depending directly on the PFC density field $\psi$, using the transformation \eqref{eq:concentration-field} as
\begin{equation}\label{eq:viscosity-pfc}
\tilde{\eta}(\rb) = \tilde{\eta}(\psi(\rb))=\big(\frac{\bar{\eta}_p}{\bar{\eta}_f}-1\big)\phi(\psi),
\end{equation}

% ---------------------------------------------------------------------------------------------------
\subsection{Peak velocities}
To approximate the particle velocities $\mathbf{v}_i$ we follow the approach of Rauscher \cite{Rauscher2010} and consider for the derivation a curl-free velocity field. Let $\mathbf{u}$ be given with the property $\nabla\times\mathbf{u}=0$ and $\nabla\cdot\mathbf{u}=0$. Then there exist a potential field $\Psi$ such that
\begin{equation}\label{eq:potential_field}
\mathbf{u} = -\frac{1}{\gamma m}\nabla\Psi,\quad\Delta\Psi = 0.
\end{equation}
Following the argumentation of \cite{Rauscher2010} the flow potential $\Psi$ acts as an external potential that drives the particle density. In DDFT models this external potential enters the free-energy, by $\mathcal{F}^\ast[\varrho]:=\mathcal{F}_\text{H}[\varrho] + \mathcal{F}_\text{ext}[\varrho]$, with
\[\mathcal{F}_\text{ext}[\varrho] = \int\Psi(\rb)\varrho\,\dr\]
Inserting $\mathcal{F}^\ast$ into \eqref{eq:archer_velocity} instead of $\mathcal{F}_\text{H}$ leads to
\begin{align*}
m\varrho\left(\partial_t\mathbf{v} + (\mathbf{v}\cdot\nabla)\mathbf{v} + \gamma\mathbf{v}\right) &= -\varrho\nabla\frac{\delta\mathcal{F}^\ast[\varrho]}{\delta\varrho} + \eta\Delta\mathbf{v} \\
&= -\varrho\nabla\frac{\delta\mathcal{F}_\text{H}[\varrho]}{\delta\varrho} - \varrho\nabla\Psi + \eta\Delta\mathbf{v} \\
&= -\varrho\nabla\frac{\delta\mathcal{F}_\text{H}[\varrho]}{\delta\varrho} + \gamma m \varrho\mathbf{u} + \eta\Delta\mathbf{v}
\end{align*}
and finally we arrive at
\begin{equation*}\label{eq:archer_advected}
m\varrho\left(\partial_t\mathbf{v} + (\mathbf{v}\cdot\nabla)\mathbf{v} + \gamma(\mathbf{v} - \mathbf{u})\right) = -\varrho\nabla\frac{\delta\mathcal{F}_\text{H}[\varrho]}{\delta\varrho} + \eta\Delta\mathbf{v}.
\end{equation*}

Going to the dimensionless form, by introducing length- and time-scales and inserting $\mathcal{F}_\text{sh}$ for $\mathcal{F}_\text{H}$, gives
\begin{align}
\partial_t \psi + \nabla\cdot\big((1.5 + \psi)\hat{\mathbf{v}}\big) = &\;0 \label{eq:archer_advected_dimensionless_psi} \\
(1.5 + \psi)\left(\partial_t\hat{\mathbf{v}} + (\hat{\mathbf{v}}\cdot\nabla)\hat{\mathbf{v}} + \Gamma(\hat{\mathbf{v}} - \hat{\mathbf{u}})\right) = &\frac{1}{Re}\Delta\hat{\mathbf{v}} \notag \\
-\frac{1}{Pe}(1.5 + \psi)\nabla&\frac{\delta\mathcal{F}_\text{sh}[\psi]}{\delta\psi}. \label{eq:archer_advected_dimensionless_v}
\end{align}
In the overdamped limit, $\Gamma \gg 1$, the velocity equation \eqref{eq:archer_advected_dimensionless_v} reduces to a simple expression for the velocity $\hat{\mathbf{v}}$:
\begin{equation}\label{eq:velocity_expression}
\Gamma(\hat{\mathbf{v}} - \hat{\mathbf{u}}) \simeq -\frac{1}{Pe}\nabla\frac{\delta\mathcal{F}_\text{sh}[\psi]}{\delta\psi}.
\end{equation}
Inserting this into \eqref{eq:archer_advected_dimensionless_psi} gives the advected PFC equation introduced in \cite{Praetorius2011} and considered in the context of DDFT in \cite{Rauscher2007}:
\begin{align}
\partial_t \psi + \hat{\mathbf{u}}\cdot\nabla\psi &= \frac{1}{\Gamma Pe}\nabla\cdot\left((1.5+\psi)\nabla\frac{\delta\mathcal{F}_\text{sh}[\psi]}{\delta\psi}\right) \notag \\
 &= \nabla\cdot\left(M(\psi)\nabla\frac{\delta\mathcal{F}_\text{sh}[\psi]}{\delta\psi}\right),\label{eq:density_evolution}
\end{align}
with a mobility function $M(\psi) = \frac{1}{\Gamma Pe}(1.5+\psi)$.

Though this equation can only be derived for potential flows we will use it as an approximate model for non-potential flows as well. With \eqref{eq:velocity_expression} we have found an explicit expression for the mean velocity of the particles, that can be used to formulate the forcing term \eqref{eq:force-term2} in the continuous 
form
\begin{align}\label{eq:force-term3}
\mathbf{F}^{[\text{ml}]}(\rb) &=\omega\sum_i(\hat{\mathbf{v}}_i - \hat{\mathbf{u}}(\mathbf{r}))\delta(\mathbf{r} - \mathbf{r}_i) \nonumber \\
&\approx -\frac{\omega}{\Gamma Pe}\nabla\frac{\delta\mathcal{F}_\text{sh}[\psi]}{\delta\psi}\sum_i\delta(\mathbf{r} - \mathbf{r}_i) \nonumber \\
&\approx-\frac{\omega}{\Gamma Pe} \nabla\frac{\delta\mathcal{F}_\text{sh}[\psi]}{\delta\psi}\delta_{(k)},
\end{align}
with $\delta_{(k)}$ the nascent delta function \eqref{eq:nascent-delta-function} approximating $\delta_\Omega = \sum_i\delta(\rb-\rb_i)$. The first-order approximation of this force, with \[\delta_{(0)}\approx N_0\psi_{(0)}=\frac{q_0}{\pi}\psi_{(0)}=\frac{q_0}{2\pi}\big(1+\frac{1}{A}(\psi - \bar{\psi})\big),\] thus reads 
\begin{equation}\label{eq:force-term-final}
\mathbf{F}^{[\text{ml}]}_{(0)}(\rb)=-(M_0 + M_1\psi)\nabla\psi^\natural(\rb),
\end{equation}
with $M_0=(1 - \bar{\psi}/A)\frac{\omega}{\Gamma Pe}\frac{q_0}{2\pi}$ and $M_1 = \frac{\omega}{\Gamma Pe}\frac{q_0}{2\pi A}$ and $\psi^\natural :=  \frac{\delta\mathcal{F}_\text{sh}[\psi]}{\delta\psi}$, which gives the considered fully continuous description. For $\omega\sim\Gamma$ we have $M_1 = \mathcal{O}(\frac{1}{Pe})$.

% ---------------------------------------------------------------------------------------------------
\subsection{Individual particles and number of particles}
In order to allow for particles that move freely, we add a modification introduced by \cite{Chan2009}. The authors have argued, that by limiting the field $\psi$ from below, the particle interaction can be modified. Therefore they have introduced the constraint $\psi\geq 0$, which allows to control the volume fraction of particles in the domain by changing the mean density of the system.

To implement the constraint the free energy is modifed by including a penalty term, i.e. $\mathcal{F}_\text{vpfc} := \mathcal{F}_\text{sh} + \mathcal{F}_\text{penalty}$, with
\[\mathcal{F}_\text{penalty}[\psi]=\int \beta(|\psi|^n-\psi^n)\,\text{d}\hat{\mathbf{r}},\]
with $\beta\gg 1$ and  $n$ an odd integer exponent.

The variational derivative of $\mathcal{F}_\text{penalty}$ can be found to be
\begin{equation}\label{eq:penalty_term}
b(\psi) := \frac{\delta\mathcal{F}_\text{penalty}[\psi]}{\delta\psi} = n\beta\psi^{n-1}(\operatorname{sign}(\psi)-1),
\end{equation}
with 
$\operatorname{sign}(\psi)=\left\{\begin{array}{rl}
1 & \text{for }\psi>0 \\
0 & \text{for }\psi=0 \\
-1 & \text{for }\psi<0.\end{array}\right.$

While localized states are observed also in the original PFC model for a small range of parameters in the coexistence regime \cite{Thiele2013}, we consider the approach in \cite{Chan2009, Berry2011, Robbins2012} using the penalty term \eqref{eq:penalty_term}. Here the number of particles can be controlled by choosing the mean density and the area the particles occupy. The initial density field for a collection of $N$ particles located at the positions $\mathbf{r}_i$, $i=1,\ldots,N$, is a composition of local density peaks
\[\psi_0^{(i)}(\rb) \!=\! \left\{\!\!\begin{array}{ll}
A\cdot\big(\cos(\frac{\sqrt{3}q_0}{2}\|\mathbf{r} - \mathbf{r}_i\|) + 1\big) & \text{for }\|\mathbf{r} - \mathbf{r}_i\| < \frac{d}{2} \\
0 & \text{otherwise,}
\end{array}\right.\]
summed up to
$\psi(\mathbf{r}) = \sum_{i=1}^N\psi_0^{(i)}(\mathbf{r})$.

Thus each particle occupies an area of approximately $B_p:= \pi(d / 2)^2$ in 2D. Based on the ideas in \cite{Chan2009} we set the mean density in the particles domain to $\psi_1 = \sqrt{(-48-56r)/133}$ as well as $r=-0.9$ and $q_0=1$. The last two parameters define the mean density of the system as
\[\bar{\psi} = \frac{N\cdot B_p}{B_0}\psi_1\]
with $B_0 = |\Omega|$ the area of the computational domain $\Omega$, and the parameter for the density scaling is $A=\psi_1$.

% ---------------------------------------------------------------------------------------------------
\subsection{Navier-Stokes-PFC model}
Combining all the ingredients, i.e. the Navier-Stokes equation for the solvent \eqref{eq:navier-stokes} with viscosity $\eta$ given by $\eta(\psi)$ in \eqref{eq:viscosity-pfc}, and volume force $\mathbf{F}$ by expression \eqref{eq:force-term-final}, combined with the density evolution \eqref{eq:density_evolution} with $\mathcal{F}_\text{sh}$ or $\mathcal{F}_\text{vpfc}$, gives the fully continuous Navier-Stokes PFC (NS-PFC) model
\begin{equation}\label{eq:navier-stokes-pfc-model}\begin{split}
\partial_t\hat{\mathbf{u}} + (\hat{\mathbf{u}}\cdot\nabla)\hat{\mathbf{u}} &= \nabla\cdot {\tilde{\boldsymbol{\sigma}}} - M_1\psi\nabla\psi^\natural \\
\nabla\cdot\hat{\mathbf{u}} &= 0 \\
\partial_t\psi + \hat{\mathbf{u}}\cdot\nabla\psi &= \nabla\cdot \big(M(\psi)\nabla\psi^\natural\big) \\
\psi^\natural & = \frac{\delta\mathcal{F}_\text{sh/vpfc}[\psi]}{\delta\psi}
\end{split}\end{equation}
with 
\begin{equation*}
\begin{split}
{\tilde{\boldsymbol{\sigma}}} &= -\tilde{p} {\boldsymbol{I}} + \frac{1}{Re_f}\big(1+\tilde{\eta}(\psi)\big)\big(\nabla \hat{\mathbf{u}} + \nabla \hat{\mathbf{u}}^\top\big) \\
\frac{\delta\mathcal{F}_\text{sh}[\psi]}{\delta\psi} &= \psi^3 + (r + (1+\Delta)^2)\psi \\
\frac{\delta\mathcal{F}_\text{vpfc}[\psi]}{\delta\psi} &= \psi^3 + (r + (1+\Delta)^2)\psi + b(\psi)
\end{split}
\end{equation*} 
and $\tilde{p} = \hat{p} + M_0\psi^\natural$ a rescaled pressure. Besides the definition of $\psi^\natural$, these equations have exactly the form of ``Model H'' as considered in \cite{Jacqmin1999}. In Appendix \ref{app:b} we demonstrate thermodynamic consistency of the derived model.

% ---------------------------------------------------------------------------------------------------
\section{Numerical studies}
We now turn to quantitative properties of the model and compare it with the original PFC model and the FPD approach of \cite{Tanaka2000} for various situations. We rewrite the NS-PFC system as a system of second order equations. Therefore the variational derivatives are implemented as
\begin{align*}
\frac{\delta\mathcal{F}_\text{sh}[\psi]}{\delta\psi} &= \psi^3 + (r + 1)\psi + 2\Delta\psi + \Delta\nu \\
\frac{\delta\mathcal{F}_\text{vpfc}[\psi]}{\delta\psi} &= \psi^3 + (r + 1)\psi + 2\Delta\psi + \Delta\nu + b(\psi) \\
\nu &= \Delta\psi.
\end{align*}
The system \eqref{eq:navier-stokes-pfc-model} has to be solved for $\mathbf{u}, \tilde{p}, \psi, \psi^\natural$ and $\nu$ in a domain $\Omega$ with boundary conditions depending on the concrete example. To numerically solve this system of partial differential equation we apply here an operator splitting approach \cite{Axelsson2011} with a sequential splitting, where we solve the PFC equations first, followed by the Navier-Stokes equations. In time we use a semi-implicit backward Euler discretization with a linearization of all nonlinear terms, i.e. a one-step Newton iteration. In space we discretize using a finite element method, with Lagrange elements, e.g. a $P^2/P^1$ Taylor-Hood element for the Navier-Stokes equation and a $P^2$ element for $\psi$, $\psi^\natural$ and $\nu$ in the PFC equation. We further use adaptive mesh refinement, leading to an enhanced resolution along the particles. The system is solved using the parallel adaptive finite element framework AMDiS \cite{AMDiS,AMDiS2014}. 

% ---------------------------------------------------------------------------------------------------
\subsection{Crystallization}
The first numerical examples uses $\mathcal{F}_\text{sh}$, and considers crystallization processes in flowing environments. The fluid is driven by boundary conditions. In the first case we consider a rotating fluid, i.e. a gyre flow, and in the second case a Poiseuille flow with a parabolic inflow velocity profile.

\subsubsection{Rotating crystals}
A crystal grain is places in a rotating fluid initially given by
\begin{equation}\label{eq:gyre_flow}
\partial_t\rb = \mathbf{u}_0(x,y) = \left(\begin{array}{rr}
C\sin(\pi \frac{x}{\text{dim}_x})\cos(\pi \frac{y}{\text{dim}_y}) \\
-C\cos(\pi \frac{x}{\text{dim}_x})\sin(\pi \frac{y}{\text{dim}_y})
\end{array}\right)
\end{equation}
in a domain $(x,y)\in\Omega = [0,\text{dim}_x]\times[0,\text{dim}_y]$. For the numerical experiment we have chosen $\text{dim}_x=\text{dim}_y=42 d$. The boundary conditions for the Navier-Stokes equations are set by $\mathbf{u}_0$.

We start the growth process with an initial grain of radius $2d$ in an undercooled environment with parameters $r=-0.3$ and mean density $\bar{\psi}=-0.35$. The mobility function is set to $M(\psi)=\psi+1.5$ and the force scaling to $M_1=1$. The fluid Reynolds number is set to $Re_f=1$ and the viscosity ratio to $\bar{\eta}_p / \bar{\eta}_f = 100$. For the concentration field that defines the profile of the viscosity, we have used an approximation of $\psi_{(0)}$, i.e.
\[\phi:=\psi_{(0)}\approx \frac{\psi - \min_{\Omega}(\psi)}{\max_{\Omega}(\psi) - \min_{\Omega}(\psi)}.\]
Thus the fluid viscosity is high in particles, low in between particles and takes an intermediate value in the isotropic phase away from the crystal. 

\begin{figure}
\begin{center}
\includegraphics{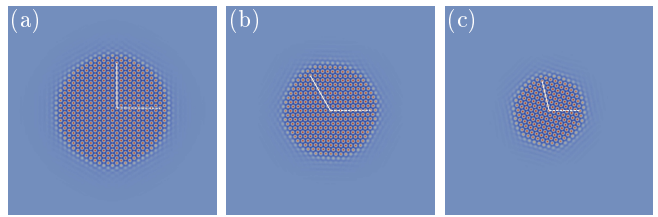}
\caption{(Color online) Final growth-shapes of the crystal in a flowing environment at time $t=3000$. Shown is the particle density $\psi$ with color red corresponds to high density and blue to low density. The fluid velocity denoted by $C$: (a) $C=0$, (b) $C=0.5$, (c) $C=1$. The white angles show the crystal orientation and thus give an indication for crystal rotation.}\label{fig:crystal_growth}
\end{center}
\end{figure}

In Fig. \ref{fig:crystal_growth} the growth shapes for different velocities $C$ are shown at the same simulation time. For a still fluid ($C = 0$), i.e. no advection, the final shape is the largest and the size of the crystal decreases for increasing velocity. For the largest considered velocity $C=1$ also the faceting of the crystal is more pronounced than for the case of no induced fluid flow. The stationary images show also that the crystal rotates during the growth process. This can be seen at the different crystal orientations in (a), (b) and (c) indicated by the white angle. 

\begin{figure}
\begin{center}
\includegraphics{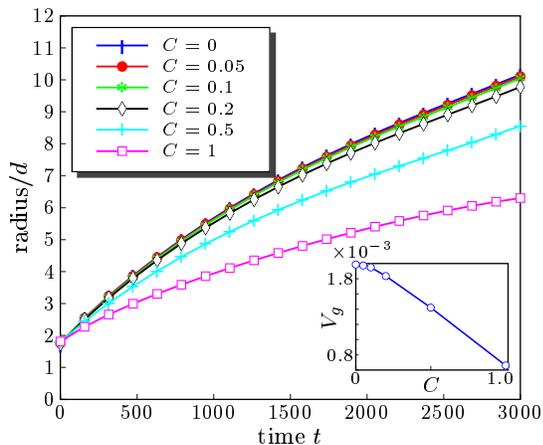}
\caption{Radius of the crystal divided by lattice constant over dimensionless time, for various fluid velocities $C$. In the inlet the growth velocity of the crystal normalized by the lattice constant $V_g$ is shown for the final time $t = 3000$.}\label{fig:crystal_growth_radius}
\end{center}
\end{figure}

The growth process is analyzed in Fig. \ref{fig:crystal_growth_radius}, showing the radius of the growing crystal over simulation time. The growth velocity strongly depends on the induced fluid velocity, as shown in the inlet plot of Fig. \ref{fig:crystal_growth_radius}. The crystal grows slower for larger induced fluid velocity. So one direction of the coupling is cleary shown, the fluid influences the crystallization.

\begin{figure}
\begin{center}
\includegraphics{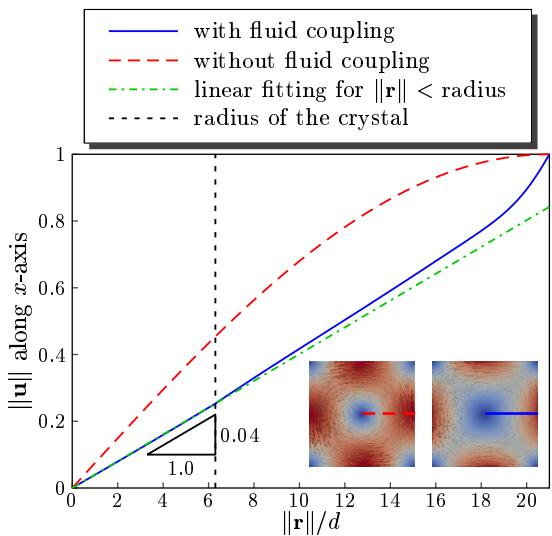}
\caption{(Color online) Fluid velocity at time $t=3000$ extracted from positive $x$-axis, as indicated by the lines in the inlet pictures. The slope $0.04$ corresponds to the angular velocity of the fluid in the region with radius less than the crystal radius. In the inlet the magnitude of fluid velocity in the domain $\Omega$ for $C=1$ is shown. Left: fluid flow not influenced by the crystal, Right: crystal slows down the fluid due to higher viscosity in the region of particles.}\label{fig:crystal_growth_fluid_velocity_x}
\end{center}
\end{figure}

Also the opposite can be found. The crystal also changes the velocity profile of the fluid. In Fig. \ref{fig:crystal_growth_fluid_velocity} the velocity profiles of two fluids are compared. The left shows the profile of a fluid with no backcoupling of the density field to the Navier-Stokes equations. This essentially just shows the initial profile $\mathbf{u}_0$. The rigt shows the velocity profile for the full NS-PFC model with $C = 1$. We observe different magnitudes of the velocity, whereas the streamlines do not change qualitatively. A more detailed analyses of the velocity profile along the $x$-axis from the center to the boundary of the domain can be found in Fig. \ref{fig:crystal_growth_fluid_velocity_x}. With fluid coupling a linear increase of the magnitude in the domain of the crystal, indicated by the black dashed line, is observed, which is lower than the presribed initial profile. The crystal acts as a rotating solid in the fluid, with normalized angular velocity $\omega=\|\mathbf{u}\| / (\|\rb\|/d)=0.04$. Away from the crystal the velocity increases up to the prescribed boundary velocity $C$.

\subsubsection{Translating crystals}
In the second case the crystals grow in a Poiseuille flow. In a narrow channel we enforce a parabolic velocity velocity at the inflow boundary, i.e. \[\mathbf{u}_0(x,y) = \left(4C \bar{y}(1 - \bar{y}), 0\right)^\top,\; \bar{y} := \frac{y}{\text{dim}_y},\] with maximal inflow velocity $C$ and top/bottom boundary velocity set to zero. Again we start with an initial grain of radius $2d$ in the center of a box $\Omega$ with dimensions $\text{dim}_x=168d$ and $\text{dim}_y=42d$. The simulation parameters are the same as above in the case of a rotating fluid.

\begin{figure}
\begin{center}
\includegraphics{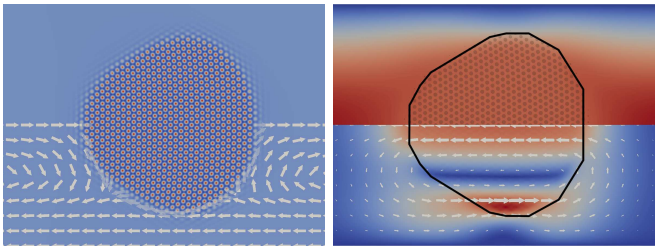}
\caption{(Color online) Crystal shape and corresponding velocity profile at time t=3800 in a narrow channel. Left: density field $\psi$, with arrows in the lower half corresponding to $\mathbf{u} - \mathbf{v}_\text{crystal}$, i.e. fluid velocity relative to the translation velocity of the crystal. Right: fluid velocity $\mathbf{u}$ with contour line that indicates the shape of the crystal. In the lower half the velocity relative to the channel flow velocity, i.e. $\mathbf{u}-\mathbf{u}_0$, is shown. Color red corresponds to high values and blue to low values.}\label{fig:translate_3shapes}
\end{center}
\end{figure}

The shape of the growing crystal is influenced by the fluid, which induces an anisotropy. This can be seen in Fig. \ref{fig:translate_3shapes}, where the shape corresponding to a fluid velocity $C=0.15$ is shown in a clipping of the whole domain $\Omega$. The flow is from left to right. The particle density $\psi$ is shown in the left image together with 
the velocity relative to the velocity of the translating crystal, i.e. $\mathbf{v}_\text{crystal}=(v_\text{crystal}, 0)^\top$ with $v_\text{crystal}\approx 0.124$. The right image shows on top the absolute value of the velocity, with a constant value within the crystal, and on the bottom the flow velocity relative to the initial velocity $\mathbf{u}_0$ showing an elongated vortex. In case of no fluid coupling the crystal grows isotropically to a circular shape, as in the example above. 

Thus, also for Poiseuille flow we see a coupling in both directions, the shape of the crystal is influenced by the flowing environment and the fluid velocity is influenced by the crystal.

% ---------------------------------------------------------------------------------------------------
\subsection{Sedimentation} 
In the following we apply the NS-PFC model to a collection of individual particles to show the applicability as a model for particle dynamics. We therefore consider $\mathcal{F}_\text{vpfc}$. For the penalty term \eqref{eq:penalty_term} we use the parameters $(n,\beta)=(3, 2000)$ in all of the following simulations. The Reynolds number and viscosity ratio are chosen as before, but the viscosity profile is now given by
\[\phi:=\psi_{(0)}\approx \frac{\psi}{\max_\Omega(\psi)}.\]
Thus, we have the lower fluid viscosity away from the particles and a high viscosity on the particles. In order to stabilize the shape of the particles we increase the diffusional part, i.e. the Peclet number $Pe$, respective the mobility function $M(\psi)$. We have chosen $M(\psi)\equiv 16$ in the following examples.

\subsubsection{One spherical particle in a confinement}
The objective of this study is to calculate the position and velocity of one spherical particle (circular disk) settling down in an enclosure due to a gravitational force $\mathbf{g}$. In order to include this force, we use a Bousinesq approximation and add the forcing term $\mathbf{F}_\mathbf{g}:=\phi(\psi)\mathbf{g}$ to the Navier-Stokes equations in \eqref{eq:navier-stokes-pfc-model}.

The box dimensions are chosen to be multiples of the to particle size. All lengths are again normalized by the particle interaction distance $d=4\pi / \sqrt{3}$, i.e. the lattice constant. We consider the following boundary conditions:

\begin{tabular}{lr}
\parbox{.63\linewidth}{\begin{align*}
\psi &= 0 &&\text{at }\partial\Omega_\text{l}\cup\partial\Omega_\text{r}\cup\partial\Omega_\text{b} \\
\partial_\mathbf{n}\psi &= 0 &&\text{at }\partial\Omega_\text{t} \\
\mathbf{u} &= 0 &&\text{at }\partial\Omega_\text{l}\cup\partial\Omega_\text{r}\cup\partial\Omega_\text{b} \\
\tilde{\boldsymbol{\sigma}}\cdot\mathbf{n}_\Gamma &= 0 &&\text{at }\partial\Omega_\text{t}, 
\end{align*}} & \parbox{.3\linewidth}{\includegraphics[scale=0.8]{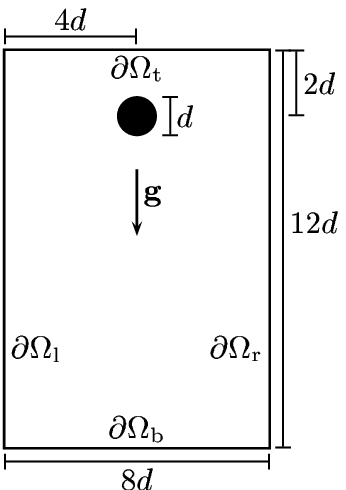}}
\end{tabular}
with $\mathbf{n}_\Gamma$ the outer normal to $\Gamma:=\partial\Omega$.

\begin{figure}
\begin{center}
\begin{tabular}{c}
\includegraphics{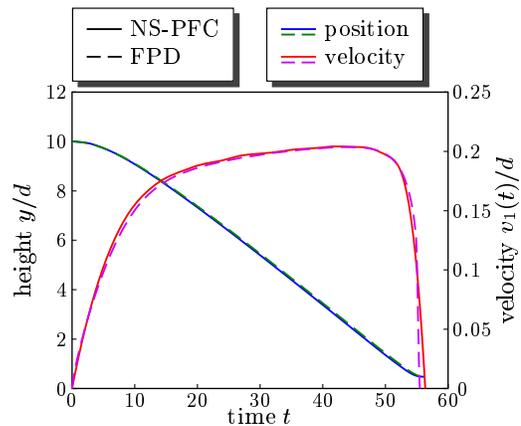}
\end{tabular}
\caption{Trajectory and velocity of one particle settling down in a box filled with a liquid with fluid viscosity $\bar{\eta}_f=0.1$, particle viscosity $\bar{\eta}_p=10$, and with gravitational force $\mathbf{g}=(0,-1)$. Left: vertical position of the particle, starting from an initial height of $10d$, Right: effective velocity of the particle, i.e. $v_1(t)^2 = ({\dot{x}(t)}^2 + {\dot{y}(t)}^2) / d^2$}
\label{fig:1particle}
\end{center}
\end{figure}

Due to the symmetry of the system we expect a symmetric trajectory, a straight line in the center of the box with the particle slowing down at the bottom. Fig. \ref{fig:1particle} shows the $y(t)$ component of the evolution curve $\big(x(t), y(t)\big)$ in comparison with FPD simulations. We further show the comparison of the velocity profiles. For both criteria we obtain an excellent agreement.

In the FPD setup we have used the normalized density field $\psi_{(0)}(\rb)$ as concentration field instead of a $\tanh$-profile and for treatment of the wall-boundary we have introduced a repulsive potential \[{\cal{V}}_B(k,p)(\rb):=k\big(d^{-1}\operatorname{dist}_{\partial\Omega}(\rb)\big)^{p},\] with $k=1, p=20$ and $\operatorname{dist}_{\partial\Omega}(\rb)$ the distance of $\rb$ to the boundary $\Gamma$ of the domain $\Omega$. 

Further care is needed in order to guarantee a symmetric solution. Within both approaches we use a symmetric triangulation of the domain and symmetric quadrature rules. Otherwise we get symmetry breaking in the trajectories, since the motion on a straight line is unstable with respect to small perturbations, as it is also pointed out in the work of \cite{Glowinski2001}.

% ---------------------------------------------------------------------------------------------------
\subsubsection{Two interacting particles}
\begin{figure}
\begin{center}
\includegraphics{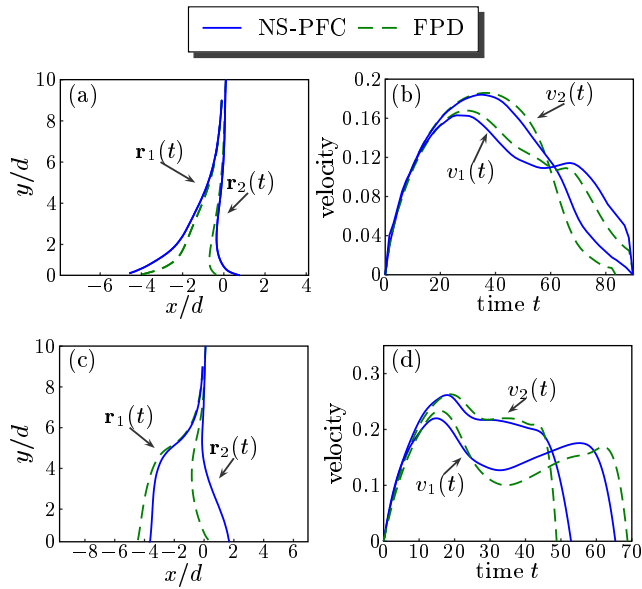}
\caption{Two particles settling down in an enclosure. Fluid viscosity is set to $\bar{\eta}_f=1$ (a,b) and $\bar{\eta}_f=0.1$ (c,d), particle viscosity to $\bar{\eta}_p=100$ (a,b) and $\bar{\eta}_p =10$ (c,d) and mobility of the NS-PFC-model to $M(\psi)\equiv 2$. Left: trajectories of the particles with coordinates $\mathbf{r}_i(t)=\big(x_i(t),y_i(t)\big)$, $i=1,2$, Right: absolute velocities: $v_i(t)^2 = ({\dot{x}_i(t)}^2 + {\dot{y}_i(t)}^2)/d^2$, $i=1,2$.}
\label{fig:2particle_nu1}
\end{center}
\end{figure}

For two particles sedimenting in a box additional hydrodynamic interactions are expected to influence the motion of the particles. We expect to see the phenomena of trailing, drafting, kissing and tumbling of the particles, as found in experimental studies \cite{Fortes2006} and also observed in several numerical studies with various methods, e.g. \cite{Hu1992, Ritz1999, Glowinski2001}. Again we compare against FPD simulations where we have to apply direct particle-particle interaction potentials, defined as ${\cal{V}}(\rb):=k\big(\big(\frac{\rb}{d}\big)^{p_1} - 2\big(\frac{\rb}{d}\big)^{p_2}\big)$, with $(k,p_1,p_2) = (1, 12, 6)$ and a boundary interaction potential ${\cal{V}}_B$ as above. Since we do not have a one-to-one mapping between these potentials and their representation in ${\cal{F}_{\text{vpfc}}}$, and since the PFC-model introduces additional diffusion due to a non-vanishing mobility function $M(\psi)$, equality of particle trajectories and particle velocities can not be expected. However, the results qualitatively agree, as can be seen in Fig. \ref{fig:2particle_nu1} for different fluid viscosities. To analyse the dependency of the trajectories on the considered interaction potential ${\cal{V}}$ FPD simulations with different potentials, i.e.different parameters in the Lennard-Jones type interaction and purely repulsive interaction are performed and compared with each other. The obtained differences in the trajectories and particle velocity are in the same order as the differences if compared with the NS-PFC simulations (results not shown).

The system considered here consists of two particles placed below each other with a small (symmetric) displacement relative to the middle vertical axis. The initial configuration is chosen as $\rb_1=(-0.1d, 9d)$ and $\rb_2=(0.1d, 10d)$, with boundary conditions as for the case of one particle. The box size is chosen wider compared to the one-particle case, i.e. a width of $18d$ instead of $8d$, to further reduce boundary effects. 

The solution can also be compared qualitatively to the results in \cite{Glowinski2001,Lin2011}, where the authors have studied the sedimentation of two hard sphere particles in a narrow enclosure in a similar setup and found similar trailing and drafting phenomena.
However, they are not as strong as in the FPD or our simulations. The particles start in nearly contact and accelerate up to a critical time, when they start moving apart from each other. In the visualized scenarios in Fig. \ref{fig:2particle_nu1} the particle behind overtakes the other one and reaches the bottom first. Compared with FPD in our simulations the particles move further apart from each other and the velocity decreases in a similar way up to contact with the lower boundary. 

% ---------------------------------------------------------------------------------------------------
\subsubsection{Many particles in an enclosure}
\begin{figure}
\begin{center}
\includegraphics{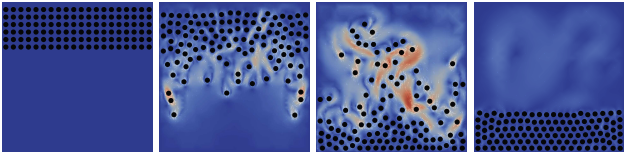}
\caption{(Color online)  Four snapshots of the sedimentation simulation for 120 particles in a square box. Color red corresponds to hight absolute velocity and blue to low velocity. Left: initial configuration of particles. Second image: an instability of the particle front, starting from the boundaries. Third image: particles start to sediment on the bottom. Right: final compressed sediment of particles.}
\label{fig:120particle}
\end{center}
\end{figure}

Already with three particles the interaction and motion of the particles becomes chaotic, as pointed out in \cite{Wolf1997} and is discussed in detail in the review \cite{Ramaswamy2001}. Therefore a direct comparison of trajectories is no longer meaningful. However, considering not only a few, but a larger number of particles in a bounded box under gravity give rise to new effects. Particles settle down not homogeneously, but their dynamics strongly depend on the distance to the walls. During the sedimentation process Rayleigh-Taylor-like instabilities and fingering occur and a compression of the particle lattice at the bottom of the box is seen. To demonstrate the possibility of our approach to deal with moderate numbers of particles we aim to observe these phenomena. We studied a situation of 120 particles arranged in a square lattice in the upper part of a square domain. The initial distance of neighboring particles is set to the lattice constant $d$. The width of the box is chosen so that 20 particles fit perfectly in one horizontal line, i.e. we have $\text{dim}_x = \text{dim}_y =20d$. Boundary conditions are similar to the case of one, respective two particles. For a gravitational force $\mathbf{g}=(0,-2)^\top$ we have simulated the sedimentation process in a fluid with viscosity ratio $\bar{\eta}_p / \bar{\eta}_f=100$, as above. The particles near the side walls start settling down first and due to their motion an upwards fluid flow in the center of the domain in induced. A visualization of the sedimentation process is shown in Fig. \ref{fig:120particle}. We have drawn black circular disks to indicate the particle positions. Four snapshots are shown, the initial and final configuration and two intermediate states, i.e. the beginning of the development of the instability and a snapshot with partially sedimented particles. 

\begin{figure}
\begin{center}
\includegraphics{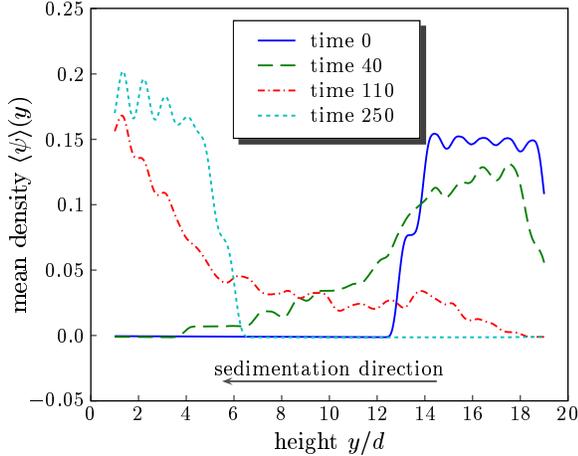}
\caption{Evolution of mean density of particles for four different time steps corresponding to the snapshots in figure \ref{fig:120particle}. The final configuration of particles in a hexagonal lattice has a higher density than the initial configuration in a square lattice.}
\label{fig:120particle_mean_density}
\end{center}
\end{figure}

In Fig. \ref{fig:120particle_mean_density} the mean particle concentration $\langle\psi\rangle(y)$ is show, which is obtained by averaging over stripes of width $d$ along the particle layers:
\[\langle\psi\rangle(y):=\int_{x=x_\text{min}}^{x_\text{max}}\int_{y'=y-0.5d}^{y+0.5d} \psi(x,y')\,\text{d}y'\,\text{d}x.\]
The high-concentration region moves from top to bottom over time and the mean particle density is higher at the bottom of the box than for the initial configuration.

%%%%%%%%%%%%%%%%%%%%%%%%%%%%%%%%%%%%%%%%%%%%%%%%%%%%%%%%%%%%%%%%%%%%%%%%%%%%%%%%%%%%%%%%%%%%%%%%%%%%%%%
\section{\label{seq:conclusion}Conclusion}
A fully continuous model is developed to simulate colloidal particles in a fluid, interacting via direct particle-particle interaction and via the induced flow fields. The method is based on ideas of dynamic density functional theory and fully resolved direct numerical simulations. The derived NS-PFC system operates on diffusive time scales and provides a qualitative approach down to the particle size. 

We have demonstrated the quality of the method in various examples, first in crystallization processes analysing the influence of a macroscopic flow field and second for three common test cases, namely the sedimentation of one, two and many particles. For one and two particles we have quantitatively compared the trajectories and velocities obtained by our simulation to simulations with the FPD method and have found good agreement. For the case of many particles we see the expected instabilities and compression at the bottom.

The formulation as a fully continuous model has several numerical advantages. We expect stable numerical behavior. For the classical PFC equation time step independent stability can be proven for the discrete scheme \cite{Backofen2007,Wise2009,Praetorius2014}. Coupling this to the Navier-Stokes equation, as considered e.g. in \cite{Aland2014} allows for larger time steps as in the explicit coupling schemes of FPD, or smoothed profile methods \cite{Nakayama2005}, and as the NS-PFC model only contains local terms, the algorithms are expected to scale independent of the number of particles. Numerical details of an efficient parallel scheme will be published elsewhere. 

\begin{acknowledgments}
The work has been funded through grant DFG Vo899/11 and FP7 IRSES 247504. We further acknowledge the provided computing resources at ZIH at TU Dresden and JSC at FZ J\"ulich.
\end{acknowledgments}

%%%%%%%%%%%%%%%%%%%%%%%%%%%%%%%%%%%%%%%%%%%%%%%%%%%%%%%%%%%%%%%%%%%%%%%%%%%%%%%%%%%%%%%%%%%%%%%%%%%%%%%
\appendix
\section{Dimensionless form} \label{app:a}

The density $\varrho$ is driven by the variational derivative of the Helmholtz free energy $\mathcal{F}_\text{H}$. This functional can be decomposed into two contributions $\mathcal{F}_\text{H} = \mathcal{F}_\text{id} + \mathcal{F}_\text{exc}$, where the ideal gas part $\mathcal{F}_\text{id}$ is known and the excess free part unknown for general systems:
\[\mathcal{F}_\text{H}[\varrho] = k_\text{B} T\int \varrho[\ln(\Lambda^d\varrho)-1]\dr + \mathcal{F}_\text{exc}[\varrho],\]
with $k_\text{B}$ Bolzmann's constant, $T$ the temperature, $\Lambda$ the thermal de-Broglie wave-length and $d$ the space dimension.

Inserting a parametrization $\varrho=\varrho(\varphi)=\bar{\varrho}(1+\varphi)$ with the density deviation $\varphi$ and reference density $\bar{\varrho}$, into the energy and expanding the ideal gas part of the energy around $\bar{\varrho}$ leads to a polynomial form of $\mathcal{F}_\text{id}$. Using a Ramakrishnan-Yussouff approximation \cite{Ramakrishnan1979} of the excess free part results in an expression of the two-point correlation function $c^{(2)}(\rb_1, \rb_2, \varrho)$:
\[\mathcal{F}_\text{exc} = C + \iint (\varrho(\rb)-\varrho_L)c^{(2)}(\rb_1, \rb_2, \varrho_L)(\varrho(\rb')-\varrho_L)\dr\dr',\]
with $\varrho_L$ a reference liquid density. This corresponds to a convolution of $(\varrho-\varrho_L)$ with $c^{(2)}$ and can thus be transformed into a product in Fourier space. Expanding $c^{(2)}$ around the wave-number zero and transforming back leads to a grandient expansion of $\mathcal{F}_\text{exc}$, that can be written in the variable $\varphi$:
\begin{align*}
\frac{1}{k_\text{B}T\bar{\varrho}}(\mathcal{F}_\text{H}[\varrho(\varphi)] - \bar{\mathcal{F}}_\text{H})&\approx\int \frac{1}{2}\varphi^2 - \frac{1}{6}\varphi^3 + \frac{1}{12}\varphi^4\dr \\
&-\int\frac{1}{2}\varphi(C_0 - C_2\Delta + C_4\Delta^2)\varphi\dr \\
&+\int D_0 + D_1\varphi\dr,
\end{align*}
with $C_0, C_2, C_4, D_0, D_1$ expansion coefficients. See e.g. \cite{Teeffelen2009, Praetorius2011} for a detailed derivation. Since we take the gradient of the variational derivative in the dynamical equations, all constant and linear terms can be neglected in the energy without changing the dynamics.

Fixing the lattice spacing $L$, the dimensionless bulk modulus of the crystal $B$ and introducing parameters $r$ and $\psi_0$, with
\begin{align*}
L^2 &:= \frac{2|C_4|}{C_2}, & B&:= \frac{C_2^2}{4|C_4|} =  \frac{\psi_0^2}{3},\\
\operatorname{sign}(C_4)&=-1, & r&:=\psi_0^{-2}\left(\frac{9}{4}-3C_0\right)-1,
\end{align*}
scaling the length by $L$, i.e. $\hat{\rb}=\hat{\rb}(\rb) := \frac{\rb}{L}$, introducing the derivatives $\hat{\nabla}:=\partial_{\hat{\rb}},\;\hat{\Delta}=\hat{\nabla}\cdot\hat{\nabla}$ and a new variable $\psi=\psi(\hat{\rb})$ as 
\[\varrho(\rb)=\bar{\varrho}(\psi_0\cdot(\psi\circ\hat{\rb})(\rb) + 1.5),\]
with $(\psi\circ\hat{\rb})(\rb) = \psi(\hat{\rb}(\rb))$, where $\circ$ acts as a function composition operator, results in the classical PFC energy
\begin{multline*}
\frac{3}{k_\text{B}T\bar{\varrho}\psi_0^4}(\mathcal{F}_\text{H}[\varrho(\psi)] - \bar{\mathcal{F}}_\text{H}) \\
\approx L^d\int \frac{1}{2}(1+r)\psi^2 + \frac{1}{4}\psi^4 +\psi\hat{\Delta}\psi + \frac{1}{2}\psi\hat{\Delta}^2\psi\,\text{d}\hat{\mathbf{r}} \\
=: L^d\mathcal{F}_\text{sh}[\psi].
\end{multline*}

We consider the variational derivative of $\mathcal{F}_\text{H}$ and relate it to the variational derivative of $\mathcal{F}_\text{sh}$:
\begin{align*}
\frac{\delta \mathcal{F}_\text{H}[\varrho]}{\delta\varrho} &= \frac{1}{L^d}\left(\frac{\delta\mathcal{F}_\text{H}[\varrho\circ\rb]}{\delta(\varrho\circ\rb)}\circ\hat{\rb}\right) \\
&\approx \frac{1}{L^d}\left(\frac{\delta \big(\frac{1}{3}k_\text{B} T\bar{\varrho}\psi_0^4 L^d \mathcal{F}_\text{sh}[\psi] + \bar{\mathcal{F}}_\text{H}\big)}{\delta(\varrho\circ\rb)}\circ\hat{\rb}\right) \\
&= \frac{1}{3}k_\text{B} T\bar{\varrho}\psi_0^4\left(\frac{\delta \mathcal{F}_\text{sh}[\psi]}{\delta\psi}\frac{\delta\psi}{\delta(\varrho\circ\rb)}\circ\hat{\rb}\right) \\
&=\frac{1}{3}k_\text{B} T\psi_0^3\left(\frac{\delta \mathcal{F}_\text{sh}[\psi]}{\delta\psi}\circ\hat{\rb}\right).
\end{align*}
Inserting the parametrization of $\varrho$ into the dynamical equations \eqref{eq:continuity_equation} and \eqref{eq:archer_velocity}, fixing $\psi_0=1$ for simplicity and using the length scaling $\hat{\rb}$ gives
\begin{align*}
(1.5+\psi)\big(\partial_t\mathbf{v} + \frac{1}{L}(\mathbf{v}\cdot\hat{\nabla})\mathbf{v} + \gamma\mathbf{v}\big) &= \frac{k_\text{B} T}{3m L}\hat{\nabla}\frac{\delta \mathcal{F}_\text{sh}[\psi]}{\delta\psi}\\
&\quad+ \frac{\eta}{m\bar{\varrho}L^2}\hat{\Delta}\mathbf{v} \\
\partial_t\psi = -\frac{1}{L}\hat{\nabla}\cdot\big((1.5+\psi)\mathbf{v}\big).
\end{align*}
Introducing the dimensionless variables $\hat{t}:=t V_0/L$ and $\hat{\mathbf{v}}:=\mathbf{v}/V_0$ finally gives the dimensionless dynamical equations
\begin{align*}
(1.5+\psi)\big(\partial_{\hat{t}}\hat{\mathbf{v}} + (\hat{\mathbf{v}}\cdot\hat{\nabla})\hat{\mathbf{v}} + \frac{\gamma L}{V_0}\hat{\mathbf{v}}\big) &= \frac{k_\text{B} T}{3m V_0^2}\hat{\nabla}\frac{\delta \mathcal{F}_\text{sh}}{\delta\psi} \\
&\quad+ \frac{\eta}{m\bar{\varrho}L V_0}\hat{\Delta}\hat{\mathbf{v}} \\
\partial_{\hat{t}}\psi = -\hat{\nabla}\cdot\big((1.5+\psi)\hat{\mathbf{v}}\big).
\end{align*}
By defining the dimensionless numbers \[Pe = \frac{3m V_0^2}{k_\text{B} T},\; Re = \frac{m\bar{\varrho} L V_0}{\eta}\;\text{ and }\;\Gamma=\frac{\gamma L}{V_0}\] as above, we find equations \eqref{eq:archer_velocity_non-dimensionlized}-\eqref{eq:archer_velocity_non-dimensionlized2}, where we have neglected the hat symbol on the derivatives for readability.

\section{Energy dissipation} \label{app:b}
To demonstrate thermodynamic consistency of the model we assume that the total energy of the system is composed of the Helmholtz-free energy $\mathcal{F}_\text{H}$, respective an appropriate approximation of this functional, and the kinetic energy \[\mathcal{F}_\text{kin} = \frac{\rho_f}{2} \int \|\mathbf{u}\|^2\,\dr\] 
of the surrounding fluid. To be consistent with the dynamic equations \eqref{eq:navier-stokes-pfc-model} we focus on the dimensionless energies by introducing length- and time-scales as above and by defining dimensionless variables denoted by a hat symbol. Additionally we normalize the energies:
\[\hat{\mathbf{r}} = \mathbf{r}/L,\; \hat{t} = t V_0 / L,\; \hat{\mathbf{u}} = \mathbf{u}/V_0,\; \hat{\mathcal{F}}_* = \mathcal{F}_* / (V_0 L^2 \bar{\eta}_f).\]
This gives us the dimensionless kinetic energy
\[\hat{\mathcal{F}}_\text{kin} = \frac{Re_f}{2}\int \|\hat{\mathbf{u}}\|^2\,\text{d}\hat{\mathbf{r}}\]
and by considering the correct scaling of the Swift-Hohenberg energy $\mathcal{F}_\text{H}\approx k_B T\bar{\varrho}\frac{L^3}{3} \mathcal{F}_\text{sh}$ (see appendix \ref{app:a}) we find
\[\hat{\mathcal{F}}_H = \frac{1}{Sc}\int \frac{1}{4}\psi^4 + \frac{1}{2}\psi\big(r + (q_0^2 + \Delta)^2\big)\psi\,\text{d}\hat{\mathbf{r}},\]
with the Schmidt number $Sc$ given by
\[Sc = \frac{Pe}{Re_f} \rrho,\quad\text{ with }\rrho:=\frac{\rho_f}{m\bar{\varrho}}.\]

The total dimensionless energy to be considered now reads
\[\hat{\mathcal{F}}_\text{tot} = \hat{\mathcal{F}}_\text{kin} + \hat{\mathcal{F}}_\text{H}.\]
In the following we consider only nondimensional variables and for readability drop the hat symbols.

We assume that the evolution equations for momentum  and mass conservation read
\begin{equation}\label{eq:balance-equations}\begin{split}
\dt\mathbf{u}+(\mathbf{u}\cdot\nabla)\mathbf{u} &= \nabla\cdot \tilde{\boldsymbol{\sigma}} + \mathbf{F}\\
\nabla\cdot\mathbf{u} &= 0\\
\dt\psi+\mathbf{u}\cdot\nabla\psi &= -\nabla\cdot\mathbf{j}
\end{split}\end{equation}
where the volume force $\mathbf{F}$ and the flux $\mathbf{j}$ need to be determined to justify thermodynamic consistency. Let $\Omega$ be a fixed domain with Lipschitz-boundary $\Gamma$. The time-evolution of the energy $\dot{\mathcal{F}}_\text{tot}$ can be split into
\begin{align}
\dot{\mathcal{F}}_\text{kin} &= Re_f\int_\Omega \mathbf{u}\cdot\dt\mathbf{u}\,\dr \notag \\
&= Re_f\int_\Omega \mathbf{u}\cdot(-(\mathbf{u}\cdot\nabla)\mathbf{u} + \nabla\cdot\tilde{\boldsymbol{\sigma}} + \mathbf{F})\,\dr \notag \\
\dot{\mathcal{F}}_\text{H} &= \frac{1}{Sc}\int_\Omega \frac{\delta\mathcal{F}_\text{sh}[\psi]}{\delta\psi} \dt\psi\,\dr. \label{eq:energy-evolution}
\end{align}
Using incompressibility and integration by parts and the relations
\begin{align*}
\frac{1}{2}\nabla\big(\|\mathbf{u}\|^2\big) &= (\mathbf{u}\cdot\nabla)\mathbf{u} - (\nabla\times\mathbf{u})\times\mathbf{u} \\
\big(f\,\nabla\mathbf{u},\, \mathbf{D}(\mathbf{u})\big)_\Omega &= \big(f\,\mathbf{D}(\mathbf{u}),\, \mathbf{D}(\mathbf{u})\big)_\Omega
\end{align*}
for a scalar field $f=f(\mathbf{r})$ and the inner product $(\mathbf{A}, \mathbf{B})_\Omega = \int_\Omega \mathbf{A}:\mathbf{B}\,\text{d}\mathbf{r}$, we get 
\begin{align*}
\int_\Omega \mathbf{u}\cdot(\mathbf{u}\cdot\nabla)\mathbf{u} &= \int_\Omega \frac{1}{2}\mathbf{u}\cdot\nabla\big(\|\mathbf{u}\|^2\big) + \mathbf{u}\cdot\big[(\nabla\times\mathbf{u})\times\mathbf{u}\big]\,\text{d}\mathbf{r} \\
&= \frac{1}{2}\int_\Gamma (\mathbf{u}\cdot\mathbf{n}_\Gamma)\|\mathbf{u}\|^2\,\text{d}\Gamma \\
&= 0\;\text{ for }\left\{\begin{array}{rl}
\mathbf{u}\cdot\mathbf{n}_\Gamma = 0 & \text{(no-penetration)} \\
\mathbf{u} = 0 & \text{(no-slip)},
\end{array}\right.
\end{align*}
\begin{align*}
\int_\Omega \mathbf{u}\cdot\nabla\cdot\tilde{\boldsymbol{\sigma}}\,\text{d}\mathbf{r} &= \underbrace{\int_\Omega -\nabla\mathbf{u}:\tilde{\boldsymbol{\sigma}}\,\text{d}\mathbf{r}}_{\leq 0} + \underbrace{\int_\Gamma \mathbf{u}\cdot\tilde{\boldsymbol{\sigma}}\cdot\mathbf{n}_\Gamma\,\text{d}\Gamma}_{(\ast)} \\
(\ast) &= 0\;\text{ for }\left\{\begin{array}{rl}
\tilde{\boldsymbol{\sigma}}\cdot\mathbf{n}_\Gamma = 0 & \text{(no-flux)} \\
\mathbf{u} = 0 & \text{(no-slip)}.
\end{array}\right.
\end{align*}
Thus we get for the kinetic part of the energy, in case of no-slip boundary conditions, the estimate
\[\dot{\mathcal{F}}_\text{kin} \leq Re_f\int_\Omega\mathbf{u} \cdot \mathbf{F} \, \dr.\]
The derivative of the PFC-part of the energy evolution reads
\[\dot{\mathcal{F}}_\text{H} = \frac{1}{Sc}\int_\Omega \mathbf{j}\nabla\frac{\delta\mathcal{F}_\text{sh}[\psi]}{\delta\psi} - \mathbf{u}\cdot \frac{\delta\mathcal{F}_\text{sh}[\psi]}{\delta\psi}\nabla\psi\,\dr.\]
By choosing the flux $\mathbf{j}$ proportional to $-\nabla\delta\mathcal{F}_\text{sh}[\psi] / \delta\psi$, e.g. \[\mathbf{j} = -M(\psi)\nabla\frac{\delta\mathcal{F}_\text{sh}[\psi]}{\delta\psi},\] with $M(\psi)$ any positive definite function, we find for the total energy evolution
\begin{align*}
\dot{\mathcal{F}}_\text{tot} &\leq \int_\Omega \mathbf{u} \cdot \big[Re_f\mathbf{F} - \frac{1}{Sc} \frac{\delta\mathcal{F}_\text{sh}[\psi]}{\delta\psi} \nabla\psi\big]\,\dr 
\end{align*}
and can choose $\mathbf{F}$ so that this integral vanishes, i.e.
\[\mathbf{F} = \frac{1}{Re_f Sc}\frac{\delta\mathcal{F}_\text{sh}[\psi]}{\delta\psi} \nabla\psi = \frac{\rrho}{Pe}\frac{\delta\mathcal{F}_\text{sh}[\psi]}{\delta\psi} \nabla\psi.\]

Using incompressibility again, we get the relation to the force and flux terms derived before. For no-slip boundary conditions, we have
\[\int_\Omega -\frac{1}{Sc} \mathbf{u} \cdot \frac{\delta\mathcal{F}_\text{sh}[\psi]}{\delta\psi} \nabla\psi\,\dr  = \int_\Omega \frac{1}{Sc} \mathbf{u} \cdot \psi\nabla \frac{\delta\mathcal{F}_\text{sh}[\psi]}{\delta\psi}\,\dr\]
and thus the force
\begin{equation}\label{eq:force-term-energy}
\mathbf{F} = -\frac{\rrho}{Pe}\psi\nabla\frac{\delta\mathcal{F}_\text{sh}[\psi]}{\delta\psi}
\end{equation}
and with $M_1 = \rrho / Pe$ the above set of equations \eqref{eq:navier-stokes-pfc-model}. Our derived continuum model thus fulfills thermodynamic consistency.

\bibliography{references}

%Merlin.mbs v4.21 2009-07-09.
\begin{thebibliography}{10}%
\makeatletter
\providecommand \@ifxundefined [1]{%
 \ifx #1\undefined \expandafter \@firstoftwo
 \else \expandafter \@secondoftwo
\fi
}%
\providecommand \@ifnum [1]{%
 \ifnum #1\expandafter \@firstoftwo
 \else \expandafter \@secondoftwo
\fi
}%
\providecommand \enquote [1]{``#1''}%
\providecommand \bibnamefont  [1]{#1}%
\providecommand \bibfnamefont [1]{#1}%
\providecommand \citenamefont [1]{#1}%
\providecommand\href[0]{\@sanitize\@href}%
\providecommand\@href[1]{\endgroup\@@startlink{#1}\endgroup\@@href}%
\providecommand\@@href[1]{#1\@@endlink}%
\providecommand \@sanitize [0]{\begingroup\catcode`\&12\catcode`\#12\relax}%
\@ifxundefined \pdfoutput {\@firstoftwo}{%
 \@ifnum{\z@=\pdfoutput}{\@firstoftwo}{\@secondoftwo}%
}{%
 \providecommand\@@startlink[1]{\leavevmode\special{html:<a href="#1">}}%
 \providecommand\@@endlink[0]{\special{html:</a>}}%
}{%
 \providecommand\@@startlink[1]{%
  \leavevmode
  \pdfstartlink
   attr{/Border[0 0 1 ]/H/I/C[0 1 1]}%
   user{/Subtype/Link/A<</Type/Action/S/URI/URI(#1)>>}%
  \relax
 }%
 \providecommand\@@endlink[0]{\pdfendlink}%
}%
\providecommand \url  [0]{\begingroup\@sanitize \@url }%
\providecommand \@url [1]{\endgroup\@href {#1}{\urlprefix}}%
\providecommand \urlprefix [0]{URL }%
\providecommand \Eprint[0]{\href }%
\@ifxundefined \urlstyle {%
  \providecommand \doi [1]{doi:\discretionary{}{}{}#1}%
}{%
  \providecommand \doi [0]{doi:\discretionary{}{}{}\begingroup
  \urlstyle{rm}\Url }%
}%
\providecommand \doibase [0]{http://dx.doi.org/}%
\providecommand \Doi[1]{\href{\doibase#1}}%
\providecommand \bibAnnote [3]{%
  \BibitemShut{#1}%
  \begin{quotation}\noindent
    \textsc{Key:}\ #2\\\textsc{Annotation:}\ #3%
  \end{quotation}%
}%
\providecommand \bibAnnoteFile [2]{%
  \IfFileExists{#2}{\bibAnnote {#1} {#2} {\input{#2}}}{}%
}%
\providecommand \typeout [0]{\immediate \write \m@ne }%
\providecommand \selectlanguage [0]{\@gobble}%
\providecommand \bibinfo [0]{\@secondoftwo}%
\providecommand \bibfield [0]{\@secondoftwo}%
\providecommand \translation [1]{[#1]}%
\providecommand \BibitemOpen[0]{}%
\providecommand \bibitemStop [0]{}%
\providecommand \bibitemNoStop [0]{.\EOS\space}%
\providecommand \EOS [0]{\spacefactor3000\relax}%
\providecommand \BibitemShut [1]{\csname bibitem#1\endcsname}%
%</preamble>
\bibitem{Furukawa2010}%
  \BibitemOpen
  \bibfield{author}{%
  \bibinfo {author} {\bibfnamefont{A.}~\bibnamefont{Furukawa}}\ and\ \bibinfo
  {author} {\bibfnamefont{H.}~\bibnamefont{Tanaka}},\ }%
  \bibfield{journal}{%
  \Doi{10.1103/PhysRevLett.104.245702}{\bibinfo {journal} {Physical Review
  Letters}}\ }%
  \textbf{\bibinfo {volume} {104}},\ \bibinfo {pages} {245702} (\bibinfo {year}
  {2010})%
  \bibAnnoteFile{NoStop}{Furukawa2010}%
\bibitem{Matsuoka2012}%
  \BibitemOpen
  \bibfield{author}{%
  \bibinfo {author} {\bibfnamefont{Y.}~\bibnamefont{Matsuoka}}, \bibinfo
  {author} {\bibfnamefont{T.}~\bibnamefont{Fukasawa}}, \bibinfo {author}
  {\bibfnamefont{K.}~\bibnamefont{Higashitani}},\ and\ \bibinfo {author}
  {\bibfnamefont{R.}~\bibnamefont{Yamamoto}},\ }%
  \bibfield{journal}{%
  \Doi{10.1103/PhysRevE.86.051403}{\bibinfo {journal} {Physical Review E}}\ }%
  \textbf{\bibinfo {volume} {86}},\ \bibinfo {pages} {051403} (\bibinfo {year}
  {2012})%
  \bibAnnoteFile{NoStop}{Matsuoka2012}%
\bibitem{Padding2006}%
  \BibitemOpen
  \bibfield{author}{%
  \bibinfo {author} {\bibfnamefont{J.}~\bibnamefont{Padding}}\ and\ \bibinfo
  {author} {\bibfnamefont{A.}~\bibnamefont{Louis}},\ }%
  \bibfield{journal}{%
  \Doi{10.1103/PhysRevE.74.031402}{\bibinfo {journal} {Physical Review E}}\ }%
  \textbf{\bibinfo {volume} {74}},\ \bibinfo {pages} {031402} (\bibinfo {year}
  {2006})%
  \bibAnnoteFile{NoStop}{Padding2006}%
\bibitem{Brady1988}%
  \BibitemOpen
  \bibfield{author}{%
  \bibinfo {author} {\bibfnamefont{J.~F.}\ \bibnamefont{Brady}}\ and\ \bibinfo
  {author} {\bibfnamefont{G.}~\bibnamefont{Bossis}},\ }%
  \bibfield{journal}{%
  \Doi{10.1146/annurev.fl.20.010188.000551}{\bibinfo {journal} {Annual Review
  of Fluid Mechanics}}\ }%
  \textbf{\bibinfo {volume} {20}},\ \bibinfo {pages} {111} (\bibinfo {year}
  {1988})%
  \bibAnnoteFile{NoStop}{Brady1988}%
\bibitem{Marconi2000}%
  \BibitemOpen
  \bibfield{author}{%
  \bibinfo {author} {\bibfnamefont{U.~M.~B.}\ \bibnamefont{Marconi}}\ and\
  \bibinfo {author} {\bibfnamefont{P.}~\bibnamefont{Tarazona}},\ }%
  \bibfield{journal}{%
  \Doi{10.1088/0953-8984/12/8A/356}{\bibinfo {journal} {Journal of Physics:
  Condensed Matter}}\ }%
  \textbf{\bibinfo {volume} {12}},\ \bibinfo {pages} {A413} (\bibinfo {year}
  {2000})%
  \bibAnnoteFile{NoStop}{Marconi2000}%
\bibitem{Archer2009}%
  \BibitemOpen
  \bibfield{author}{%
  \bibinfo {author} {\bibfnamefont{A.~J.}\ \bibnamefont{Archer}},\ }%
  \bibfield{journal}{%
  \Doi{10.1063/1.3054633}{\bibinfo {journal} {The Journal of chemical
  physics}}\ }%
  \textbf{\bibinfo {volume} {130}},\ \bibinfo {pages} {014509} (\bibinfo {year}
  {2009})%
  \bibAnnoteFile{NoStop}{Archer2009}%
\bibitem{Rauscher2007}%
  \BibitemOpen
  \bibfield{author}{%
  \bibinfo {author} {\bibfnamefont{M.}~\bibnamefont{Rauscher}}, \bibinfo
  {author} {\bibfnamefont{A.}~\bibnamefont{Dominguez}}, \bibinfo {author}
  {\bibfnamefont{M.}~\bibnamefont{Kruger}},\ and\ \bibinfo {author}
  {\bibfnamefont{F.}~\bibnamefont{Penna}},\ }%
  \bibfield{journal}{%
  \Doi{10.1063/1.2806094}{\bibinfo {journal} {The Journal of Chemical
  Physics}}\ }%
  \textbf{\bibinfo {volume} {127}},\ \bibinfo {pages} {244906} (\bibinfo {year}
  {2007})%
  \bibAnnoteFile{NoStop}{Rauscher2007}%
\bibitem{Goddard2013}%
  \BibitemOpen
  \bibfield{author}{%
  \bibinfo {author} {\bibfnamefont{B.~D.}\ \bibnamefont{Goddard}}, \bibinfo
  {author} {\bibfnamefont{A.}~\bibnamefont{Nold}}, \bibinfo {author}
  {\bibfnamefont{N.}~\bibnamefont{Savva}}, \bibinfo {author}
  {\bibfnamefont{P.}~\bibnamefont{Yatsyshin}},\ and\ \bibinfo {author}
  {\bibfnamefont{S.}~\bibnamefont{Kalliadasis}},\ }%
  \bibfield{journal}{%
  \Doi{10.1088/0953-8984/25/3/035101}{\bibinfo {journal} {Journal of Physics:
  Condensed Matter}}\ }%
  \textbf{\bibinfo {volume} {25}},\ \bibinfo {pages} {035101} (\bibinfo {year}
  {2013})%
  \bibAnnoteFile{NoStop}{Goddard2013}%
\bibitem{Toth2014}%
  \BibitemOpen
  \bibfield{author}{%
  \bibinfo {author} {\bibfnamefont{G.~I.}\ \bibnamefont{T\'{o}th}}, \bibinfo
  {author} {\bibfnamefont{L.}~\bibnamefont{Gr\'{a}n\'{a}sy}},\ and\ \bibinfo
  {author} {\bibfnamefont{G.}~\bibnamefont{Tegze}},\ }%
  \bibfield{journal}{%
  \Doi{10.1088/0953-8984/26/5/055001}{\bibinfo {journal} {Journal of Physics:
  Condensed Matter}}\ }%
  \textbf{\bibinfo {volume} {26}},\ \bibinfo {pages} {055001} (\bibinfo {year}
  {2014})%
  \bibAnnoteFile{NoStop}{Toth2014}%
\bibitem{Swift1977}%
  \BibitemOpen
  \bibfield{author}{%
  \bibinfo {author} {\bibfnamefont{J.}~\bibnamefont{Swift}}\ and\ \bibinfo
  {author} {\bibfnamefont{P.~C.}\ \bibnamefont{Hohenberg}},\ }%
  \bibfield{journal}{%
  \Doi{10.1103/PhysRevA.15.319}{\bibinfo {journal} {Physical Review A}}\ }%
  \textbf{\bibinfo {volume} {15}},\ \bibinfo {pages} {319} (\bibinfo {year}
  {1977})%
  \bibAnnoteFile{NoStop}{Swift1977}%
\bibitem{Elder2002}%
  \BibitemOpen
  \bibfield{author}{%
  \bibinfo {author} {\bibfnamefont{K.}~\bibnamefont{Elder}}, \bibinfo {author}
  {\bibfnamefont{M.}~\bibnamefont{Katakowski}}, \bibinfo {author}
  {\bibfnamefont{M.}~\bibnamefont{Haataja}},\ and\ \bibinfo {author}
  {\bibfnamefont{M.}~\bibnamefont{Grant}},\ }%
  \bibfield{journal}{%
  \bibinfo {journal} {Physical Review Letters}\ }%
  \textbf{\bibinfo {volume} {88}},\ \bibinfo {pages} {245701} (\bibinfo {year}
  {2002})%
  \bibAnnoteFile{NoStop}{Elder2002}%
\bibitem{Teeffelen2009}%
  \BibitemOpen
  \bibfield{author}{%
  \bibinfo {author} {\bibfnamefont{S.}~\bibnamefont{van Teeffelen}}, \bibinfo
  {author} {\bibfnamefont{R.}~\bibnamefont{Backofen}}, \bibinfo {author}
  {\bibfnamefont{A.}~\bibnamefont{Voigt}},\ and\ \bibinfo {author}
  {\bibfnamefont{H.}~\bibnamefont{L\"{o}wen}},\ }%
  \bibfield{journal}{%
  \Doi{10.1103/PhysRevE.79.051404}{\bibinfo {journal} {Physical Review E}}\ }%
  \textbf{\bibinfo {volume} {79}},\ \bibinfo {pages} {051404} (\bibinfo {year}
  {2009})%
  \bibAnnoteFile{NoStop}{Teeffelen2009}%
\bibitem{Glowinski2001}%
  \BibitemOpen
  \bibfield{author}{%
  \bibinfo {author} {\bibfnamefont{R.}~\bibnamefont{Glowinski}}, \bibinfo
  {author} {\bibfnamefont{T.~W.}\ \bibnamefont{Pan}}, \bibinfo {author}
  {\bibfnamefont{T.~I.}\ \bibnamefont{Hesla}}, \bibinfo {author}
  {\bibfnamefont{D.~D.}\ \bibnamefont{Joseph}},\ and\ \bibinfo {author}
  {\bibfnamefont{J.}~\bibnamefont{P\'{e}riaux}},\ }%
  \bibfield{journal}{%
  \Doi{10.1006/jcph.2000.6542}{\bibinfo {journal} {Journal of Computational
  Physics}}\ }%
  \textbf{\bibinfo {volume} {169}},\ \bibinfo {pages} {363} (\bibinfo {year}
  {2001})%
  \bibAnnoteFile{NoStop}{Glowinski2001}%
\bibitem{Uhlmann2005}%
  \BibitemOpen
  \bibfield{author}{%
  \bibinfo {author} {\bibfnamefont{M.}~\bibnamefont{Uhlmann}},\ }%
  \bibfield{journal}{%
  \Doi{10.1016/j.jcp.2005.03.017}{\bibinfo {journal} {Journal of Computational
  Physics}}\ }%
  \textbf{\bibinfo {volume} {209}},\ \bibinfo {pages} {448} (\bibinfo {year}
  {2005})%
  \bibAnnoteFile{NoStop}{Uhlmann2005}%
\bibitem{Apte2009}%
  \BibitemOpen
  \bibfield{author}{%
  \bibinfo {author} {\bibfnamefont{S.~V.}\ \bibnamefont{Apte}}, \bibinfo
  {author} {\bibfnamefont{M.}~\bibnamefont{Martin}},\ and\ \bibinfo {author}
  {\bibfnamefont{N.~A.}\ \bibnamefont{Patankar}},\ }%
  \bibfield{journal}{%
  \bibinfo {journal} {Journal of Computational Physics}\ }%
  \textbf{\bibinfo {volume} {228}},\ \bibinfo {pages} {2712} (\bibinfo {year}
  {2009})%
  \bibAnnoteFile{NoStop}{Apte2009}%
\bibitem{Kempe2012}%
  \BibitemOpen
  \bibfield{author}{%
  \bibinfo {author} {\bibfnamefont{T.}~\bibnamefont{Kempe}}\ and\ \bibinfo
  {author} {\bibfnamefont{J.}~\bibnamefont{Fr\"{o}hlich}},\ }%
  \bibfield{journal}{%
  \Doi{10.1016/j.jcp.2012.01.021}{\bibinfo {journal} {Journal of Computational
  Physics}}\ }%
  \textbf{\bibinfo {volume} {231}},\ \bibinfo {pages} {3663} (\bibinfo {year}
  {2012})%
  \bibAnnoteFile{NoStop}{Kempe2012}%
\bibitem{Tanaka2000}%
  \BibitemOpen
  \bibfield{author}{%
  \bibinfo {author} {\bibfnamefont{H.}~\bibnamefont{Tanaka}}\ and\ \bibinfo
  {author} {\bibfnamefont{T.}~\bibnamefont{Araki}},\ }%
  \bibfield{journal}{%
  \bibinfo {journal} {Physical Review Letters}\ }%
  \textbf{\bibinfo {volume} {85}},\ \bibinfo {pages} {1338} (\bibinfo {year}
  {2000})%
  \bibAnnoteFile{NoStop}{Tanaka2000}%
\bibitem{Siggia1976}%
  \BibitemOpen
  \bibfield{author}{%
  \bibinfo {author} {\bibfnamefont{E.}~\bibnamefont{Siggia}}, \bibinfo {author}
  {\bibfnamefont{B.}~\bibnamefont{Halperin}},\ and\ \bibinfo {author}
  {\bibfnamefont{P.}~\bibnamefont{Hohenberg}},\ }%
  \bibfield{journal}{%
  \Doi{10.1103/PhysRevB.13.2110}{\bibinfo {journal} {Physical Review B}}\ }%
  \textbf{\bibinfo {volume} {13}},\ \bibinfo {pages} {2110} (\bibinfo {year}
  {1976})%
  \bibAnnoteFile{NoStop}{Siggia1976}%
\bibitem{Hohenberg1977}%
  \BibitemOpen
  \bibfield{author}{%
  \bibinfo {author} {\bibfnamefont{P.}~\bibnamefont{Hohenberg}}\ and\ \bibinfo
  {author} {\bibfnamefont{B.}~\bibnamefont{Halperin}},\ }%
  \bibfield{journal}{%
  \Doi{10.1103/RevModPhys.49.435}{\bibinfo {journal} {Reviews of Modern
  Physics}}\ }%
  \textbf{\bibinfo {volume} {49}},\ \bibinfo {pages} {435} (\bibinfo {year}
  {1977})%
  \bibAnnoteFile{NoStop}{Hohenberg1977}%
\bibitem{Nakayama2005}%
  \BibitemOpen
  \bibfield{author}{%
  \bibinfo {author} {\bibfnamefont{Y.}~\bibnamefont{Nakayama}}\ and\ \bibinfo
  {author} {\bibfnamefont{R.}~\bibnamefont{Yamamoto}},\ }%
  \bibfield{journal}{%
  \Doi{10.1103/PhysRevE.71.036707}{\bibinfo {journal} {Physical Review E}}\ }%
  \textbf{\bibinfo {volume} {71}},\ \bibinfo {pages} {036707} (\bibinfo {year}
  {2005})%
  \bibAnnoteFile{NoStop}{Nakayama2005}%
\bibitem{Menzel2013}%
  \BibitemOpen
  \bibfield{author}{%
  \bibinfo {author} {\bibfnamefont{A.~M.}\ \bibnamefont{Menzel}}\ and\ \bibinfo
  {author} {\bibfnamefont{H.}~\bibnamefont{L\"{o}wen}},\ }%
  \bibfield{journal}{%
  \bibinfo {journal} {Physical Review Letters}\ }%
  \textbf{\bibinfo {volume} {110}},\ \bibinfo {pages} {055702} (\bibinfo {year}
  {2013})%
  \bibAnnoteFile{NoStop}{Menzel2013}%
\bibitem{Praetorius2011}%
  \BibitemOpen
  \bibfield{author}{%
  \bibinfo {author} {\bibfnamefont{S.}~\bibnamefont{Praetorius}}\ and\ \bibinfo
  {author} {\bibfnamefont{A.}~\bibnamefont{Voigt}},\ }%
  \bibfield{journal}{%
  \bibinfo {journal} {Macromolecular Theory and Simulations}\ }%
  \textbf{\bibinfo {volume} {20}},\ \bibinfo {pages} {541} (\bibinfo {year}
  {2011})%
  \bibAnnoteFile{NoStop}{Praetorius2011}%
\bibitem{Chan2009}%
  \BibitemOpen
  \bibfield{author}{%
  \bibinfo {author} {\bibfnamefont{P.~Y.}\ \bibnamefont{Chan}}, \bibinfo
  {author} {\bibfnamefont{N.}~\bibnamefont{Goldenfeld}},\ and\ \bibinfo
  {author} {\bibfnamefont{J.}~\bibnamefont{Dantzig}},\ }%
  \bibfield{journal}{%
  \Doi{10.1103/PhysRevE.79.035701}{\bibinfo {journal} {Physical Review E}}\ }%
  \textbf{\bibinfo {volume} {79}},\ \bibinfo {pages} {035701} (\bibinfo {year}
  {2009})%
  \bibAnnoteFile{NoStop}{Chan2009}%
\bibitem{Berry2011}%
  \BibitemOpen
  \bibfield{author}{%
  \bibinfo {author} {\bibfnamefont{J.}~\bibnamefont{Berry}}\ and\ \bibinfo
  {author} {\bibfnamefont{M.}~\bibnamefont{Grant}},\ }%
  \bibfield{journal}{%
  \Doi{10.1103/PhysRevLett.106.175702}{\bibinfo {journal} {Phys. Rev. Lett.}}\
  }%
  \textbf{\bibinfo {volume} {106}},\ \bibinfo {pages} {175702} (\bibinfo
  {month} {Apr}\ \bibinfo {year} {2011})%
  \bibAnnoteFile{NoStop}{Berry2011}%
\bibitem{Elder2004}%
  \BibitemOpen
  \bibfield{author}{%
  \bibinfo {author} {\bibfnamefont{K.~R.}\ \bibnamefont{Elder}}\ and\ \bibinfo
  {author} {\bibfnamefont{M.}~\bibnamefont{Grant}},\ }%
  \bibfield{journal}{%
  \Doi{10.1103/PhysRevE.70.051605}{\bibinfo {journal} {Phys. Rev. E}}\ }%
  \textbf{\bibinfo {volume} {70}},\ \bibinfo {pages} {051605} (\bibinfo {year}
  {2004})%
  \bibAnnoteFile{NoStop}{Elder2004}%
\bibitem{Rauscher2010}%
  \BibitemOpen
  \bibfield{author}{%
  \bibinfo {author} {\bibfnamefont{M.}~\bibnamefont{Rauscher}},\ }%
  \bibfield{journal}{%
  \bibinfo {journal} {Journal of Physics: Condensed Matter}\ }%
  \textbf{\bibinfo {volume} {22}},\ \bibinfo {pages} {364109} (\bibinfo {year}
  {2010})%
  \bibAnnoteFile{NoStop}{Rauscher2010}%
\bibitem{Thiele2013}%
  \BibitemOpen
  \bibfield{author}{%
  \bibinfo {author} {\bibfnamefont{U.}~\bibnamefont{Thiele}}, \bibinfo {author}
  {\bibfnamefont{A.~J.}\ \bibnamefont{Archer}}, \bibinfo {author}
  {\bibfnamefont{M.~J.}\ \bibnamefont{Robbins}}, \bibinfo {author}
  {\bibfnamefont{H.}~\bibnamefont{Gomez}},\ and\ \bibinfo {author}
  {\bibfnamefont{E.}~\bibnamefont{Knobloch}},\ }%
  \bibfield{journal}{%
  \Doi{10.1103/PhysRevE.87.042915}{\bibinfo {journal} {Phys. Rev. E}}\ }%
  \textbf{\bibinfo {volume} {87}},\ \bibinfo {pages} {042915} (\bibinfo {month}
  {Apr}\ \bibinfo {year} {2013})%
  \bibAnnoteFile{NoStop}{Thiele2013}%
\bibitem{Robbins2012}%
  \BibitemOpen
  \bibfield{author}{%
  \bibinfo {author} {\bibfnamefont{M.~J.}\ \bibnamefont{Robbins}}, \bibinfo
  {author} {\bibfnamefont{A.~J.}\ \bibnamefont{Archer}}, \bibinfo {author}
  {\bibfnamefont{U.}~\bibnamefont{Thiele}},\ and\ \bibinfo {author}
  {\bibfnamefont{E.}~\bibnamefont{Knobloch}},\ }%
  \bibfield{journal}{%
  \Doi{10.1103/PhysRevE.85.061408}{\bibinfo {journal} {Phys. Rev. E}}\ }%
  \textbf{\bibinfo {volume} {85}},\ \bibinfo {pages} {061408} (\bibinfo {month}
  {Jun}\ \bibinfo {year} {2012})%
  \bibAnnoteFile{NoStop}{Robbins2012}%
\bibitem{Jacqmin1999}%
  \BibitemOpen
  \bibfield{author}{%
  \bibinfo {author} {\bibfnamefont{D.}~\bibnamefont{Jacqmin}},\ }%
  \bibfield{journal}{%
  \Doi{10.1006/jcph.1999.6332}{\bibinfo {journal} {Journal of Computational
  Physics}}\ }%
  \textbf{\bibinfo {volume} {155}},\ \bibinfo {pages} {32} (\bibinfo {year}
  {1999})%
  \bibAnnoteFile{NoStop}{Jacqmin1999}%
\bibitem{Axelsson2011}%
  \BibitemOpen
  \bibfield{author}{%
  \bibinfo {author} {\bibfnamefont{O.}~\bibnamefont{Axelsson}}\ and\ \bibinfo
  {author} {\bibfnamefont{M.}~\bibnamefont{Neytcheva}},\ }%
  \emph{\bibinfo {title} {{Operator splittings for solving nonlinear, coupled
  multiphysics problems with an application to the numerical solution of an
  interface problem}}},\ \bibinfo {type} {Tech. Rep.}\ (\bibinfo {year}
  {2011})%
  \bibAnnoteFile{NoStop}{Axelsson2011}%
\bibitem{AMDiS}%
  \BibitemOpen
  \bibfield{author}{%
  \bibinfo {author} {\bibfnamefont{S.}~\bibnamefont{Vey}}\ and\ \bibinfo
  {author} {\bibfnamefont{A.}~\bibnamefont{Voigt}},\ }%
  \bibfield{journal}{%
  \Doi{10.1007/s00791-006-0048-3}{\bibinfo {journal} {Computing and
  Visualization in Science}}\ }%
  \textbf{\bibinfo {volume} {10}},\ \bibinfo {pages} {57} (\bibinfo {year}
  {2007})%
  \bibAnnoteFile{NoStop}{AMDiS}%
\bibitem{AMDiS2014}%
  \BibitemOpen
  \bibfield{author}{%
  \bibinfo {author} {\bibfnamefont{T.}~\bibnamefont{Witkowski}}, \bibinfo
  {author} {\bibfnamefont{S.}~\bibnamefont{Ling}},\ and\ \bibinfo {author}
  {\bibfnamefont{A.}~\bibnamefont{Voigt}},\ }%
  \enquote{\bibinfo {title} {Software concepts and numerical algorithms for a
  scalable adaptive parallel finite element method},}\  (\bibinfo {year}
  {2014}),\ \bibinfo {note} {accepted for publication in Advances in
  Computational Mathematics}%
  \bibAnnoteFile{NoStop}{AMDiS2014}%
\bibitem{Fortes2006}%
  \BibitemOpen
  \bibfield{author}{%
  \bibinfo {author} {\bibfnamefont{A.~F.}\ \bibnamefont{Fortes}}, \bibinfo
  {author} {\bibfnamefont{D.~D.}\ \bibnamefont{Joseph}},\ and\ \bibinfo
  {author} {\bibfnamefont{T.~S.}\ \bibnamefont{Lundgren}},\ }%
  \bibfield{journal}{%
  \Doi{10.1017/S0022112087001046}{\bibinfo {journal} {Journal of Fluid
  Mechanics}}\ }%
  \textbf{\bibinfo {volume} {177}},\ \bibinfo {pages} {467} (\bibinfo {year}
  {2006})%
  \bibAnnoteFile{NoStop}{Fortes2006}%
\bibitem{Hu1992}%
  \BibitemOpen
  \bibfield{author}{%
  \bibinfo {author} {\bibfnamefont{H.~H.}\ \bibnamefont{Hu}}, \bibinfo {author}
  {\bibfnamefont{D.~D.}\ \bibnamefont{Joseph}},\ and\ \bibinfo {author}
  {\bibfnamefont{M.~J.}\ \bibnamefont{Crochet}},\ }%
  \bibfield{journal}{%
  \Doi{10.1007/BF00717645}{\bibinfo {journal} {Theoretical and Computational
  Fluid Dynamics}}\ }%
  \textbf{\bibinfo {volume} {3}},\ \bibinfo {pages} {285} (\bibinfo {year}
  {1992})%
  \bibAnnoteFile{NoStop}{Hu1992}%
\bibitem{Ritz1999}%
  \BibitemOpen
  \bibfield{author}{%
  \bibinfo {author} {\bibfnamefont{J.~B.}\ \bibnamefont{Ritz}}\ and\ \bibinfo
  {author} {\bibfnamefont{J.~P.}\ \bibnamefont{Caltagirone}},\ }%
  \bibfield{journal}{%
  \bibinfo {journal} {International Journal for Numerical Methods in Fluids}\
  }%
  \textbf{\bibinfo {volume} {30}},\ \bibinfo {pages} {1067} (\bibinfo {year}
  {1999})%
  \bibAnnoteFile{NoStop}{Ritz1999}%
\bibitem{Lin2011}%
  \BibitemOpen
  \bibfield{author}{%
  \bibinfo {author} {\bibfnamefont{S.}~\bibnamefont{Lin}}, \bibinfo {author}
  {\bibfnamefont{Y.}~\bibnamefont{Chin}}, \bibinfo {author}
  {\bibfnamefont{J.}~\bibnamefont{Hu}},\ and\ \bibinfo {author}
  {\bibfnamefont{Y.}~\bibnamefont{Chen}},\ }%
  \bibfield{journal}{%
  \Doi{10.1002/fld.2442}{\bibinfo {journal} {International Journal for
  Numerical Methods in Fluids}}\ }%
  \textbf{\bibinfo {volume} {67}},\ \bibinfo {pages} {1771} (\bibinfo {year}
  {2011})%
  \bibAnnoteFile{NoStop}{Lin2011}%
\bibitem{Wolf1997}%
  \BibitemOpen
  \bibfield{author}{%
  \bibinfo {author} {\bibfnamefont{D.~E.}\ \bibnamefont{Wolf}}, \bibinfo
  {author} {\bibfnamefont{J.~A.~C.}\ \bibnamefont{Gallas}},\ and\ \bibinfo
  {author} {\bibfnamefont{I.~M.}\ \bibnamefont{Ja}}\ }%
  \textbf{\bibinfo {volume} {56}},\ \bibinfo {pages} {2858} (\bibinfo {year}
  {1997})%
  \bibAnnoteFile{NoStop}{Wolf1997}%
\bibitem{Ramaswamy2001}%
  \BibitemOpen
  \bibfield{author}{%
  \bibinfo {author} {\bibfnamefont{S.}~\bibnamefont{Ramaswamy}},\ }%
  \bibfield{journal}{%
  \Doi{10.1080/00018730110050617}{\bibinfo {journal} {Advances in Physics}}\ }%
  \textbf{\bibinfo {volume} {50}},\ \bibinfo {pages} {297} (\bibinfo {year}
  {2001})%
  \bibAnnoteFile{NoStop}{Ramaswamy2001}%
\bibitem{Backofen2007}%
  \BibitemOpen
  \bibfield{author}{%
  \bibinfo {author} {\bibfnamefont{R.}~\bibnamefont{Backofen}}, \bibinfo
  {author} {\bibfnamefont{A.}~\bibnamefont{R\"{a}tz}},\ and\ \bibinfo {author}
  {\bibfnamefont{A.}~\bibnamefont{Voigt}},\ }%
  \bibfield{journal}{%
  \bibinfo {journal} {Philosophical Magazine Letters}\ }%
  \textbf{\bibinfo {volume} {87}},\ \bibinfo {pages} {813} (\bibinfo {year}
  {2007})%
  \bibAnnoteFile{NoStop}{Backofen2007}%
\bibitem{Wise2009}%
  \BibitemOpen
  \bibfield{author}{%
  \bibinfo {author} {\bibfnamefont{S.~M.}\ \bibnamefont{Wise}}, \bibinfo
  {author} {\bibfnamefont{C.}~\bibnamefont{Wang}},\ and\ \bibinfo {author}
  {\bibfnamefont{J.~S.}\ \bibnamefont{Lowengrub}},\ }%
  \bibfield{journal}{%
  \Doi{10.1137/080738143}{\bibinfo {journal} {SIAM Journal on Numerical
  Analysis}}\ }%
  \textbf{\bibinfo {volume} {47}},\ \bibinfo {pages} {2269} (\bibinfo {year}
  {2009})%
  \bibAnnoteFile{NoStop}{Wise2009}%
\bibitem{Praetorius2014}%
  \BibitemOpen
  \bibfield{author}{%
  \bibinfo {author} {\bibfnamefont{S.}~\bibnamefont{Praetorius}}\ and\ \bibinfo
  {author} {\bibfnamefont{A.}~\bibnamefont{Voigt}},\ }%
  \bibfield{journal}{%
  \bibinfo {journal} {SIAM Journal of Scientific Computing}}%
   (\bibinfo {year} {2014}),\ \bibinfo {note} {accepted for publication}%
  \bibAnnoteFile{NoStop}{Praetorius2014}%
\bibitem{Aland2014}%
  \BibitemOpen
  \bibfield{author}{%
  \bibinfo {author} {\bibfnamefont{S.}~\bibnamefont{Aland}},\ }%
  \bibfield{journal}{%
  \bibinfo {journal} {Journal of Computational Physics}\ }%
  \textbf{\bibinfo {volume} {262}},\ \bibinfo {pages} {58 } (\bibinfo {year}
  {2014})%
  \bibAnnoteFile{NoStop}{Aland2014}%
\bibitem{Ramakrishnan1979}%
  \BibitemOpen
  \bibfield{author}{%
  \bibinfo {author} {\bibfnamefont{T.}~\bibnamefont{Ramakrishnan}}\ and\
  \bibinfo {author} {\bibfnamefont{M.}~\bibnamefont{Yussouff}},\ }%
  \bibfield{journal}{%
  \Doi{10.1103/PhysRevB.19.2775}{\bibinfo {journal} {Physical Review B}}\ }%
  \textbf{\bibinfo {volume} {19}},\ \bibinfo {pages} {2775} (\bibinfo {year}
  {1979})%
  \bibAnnoteFile{NoStop}{Ramakrishnan1979}%
\end{thebibliography}%

\end{document}